\documentclass[10pt,conference]{IEEEtran}
\IEEEoverridecommandlockouts
\usepackage{hyperref}
\usepackage{graphicx}
\usepackage{caption}
\usepackage[ruled,lined,linesnumbered,vlined,algo2e]{algorithm2e} 
\usepackage{listings}
\usepackage{color}
\usepackage[table]{xcolor}
\usepackage{booktabs}
\hypersetup{hidelinks=true} 
\usepackage{multirow}
\usepackage{enumitem} 
\usepackage{balance}
\newcommand*{\rom}[1]{\uppercase\expandafter{\romannumeral #1\relax}}

\definecolor{amethyst}{rgb}{0.6, 0.4, 0.8}
\definecolor{awesome}{rgb}{1.0, 0.13, 0.32}
\definecolor{burntorange}{rgb}{0.8, 0.33, 0.0}
\usepackage{url}
\usepackage{graphicx} 
\usepackage{subcaption}
\usepackage{tcolorbox}

\usepackage[numbers,sort&compress]{natbib}

\usepackage{makecell}
\usepackage{threeparttable}
\makeatletter  
\newif\if@restonecol  
\renewcommand\footnoterule{%
	\kern-3\p@
	\hrule\@width\columnwidth
	\kern2.6\p@}

\newcommand{\tool}{{\textsc{\small StoryDroid}}\xspace}

\definecolor{grey}{rgb}{0.52,0.52,0.51}
\definecolor{codegray}{rgb}{0.5,0.5,0.5}
\definecolor{codepurple}{rgb}{0.58,0,0.82}
\definecolor{backcolour}{rgb}{0.96,0.96,0.96}
\definecolor{darkblue}{rgb}{0.0,0.0,0.55}
\definecolor{cobalt}{rgb}{0.0,0.28,0.67}
 \lstdefinestyle{mystyle}{
 backgroundcolor=\color{backcolour},  
 keywordstyle=\color{darkblue},
 commentstyle=\color{grey},
 emph={replace,setAdapter,PrefEditor,InnerClass,startActivity,SearchPanel,PartList},
 emphstyle={\color{magenta}},
 numberstyle=\tiny,
 stringstyle=\color{codepurple},
 basicstyle=\scriptsize,
 breakatwhitespace=true,         
 breaklines=true,                 
 captionpos=b,                    
 keepspaces=true,                 
 numbers=left,                    
 numbersep= 5pt,                  
 showspaces=false,                
 showstringspaces=false,
 showtabs=false,   
 xleftmargin=3.0ex,       
 tabsize=2,
}

\begin{document}
\title{{StoryDroid}: Automated Generation of Storyboard for Android Apps}

\author{\IEEEauthorblockN{
		Sen Chen\IEEEauthorrefmark{1},
		Lingling Fan\IEEEauthorrefmark{1},
		Chunyang Chen\IEEEauthorrefmark{2},
		Ting Su\IEEEauthorrefmark{3},
		Wenhe Li\IEEEauthorrefmark{4},
		Yang Liu\IEEEauthorrefmark{3},
		Lihua Xu\IEEEauthorrefmark{4}
	}
	\IEEEauthorblockA{\IEEEauthorrefmark{1}East China Normal University, China
		\IEEEauthorrefmark{2}Monash University, Australia\\
		\IEEEauthorrefmark{3}Nanyang Technological University, Singapore
		\IEEEauthorrefmark{4}New York University Shanghai, China\\
		ecnuchensen@gmail.com
	}}

\maketitle
\setlength{\textfloatsep}{1pt} 

\begin{abstract}
Mobile apps are now ubiquitous. Before developing a new app, the development team usually endeavors painstaking efforts to review many existing apps {with} similar purposes.
The review process is crucial in {the} sense that it reduces market risks and provides inspiration for app development.
However, manual exploration of hundreds of existing apps by different roles (e.g., product manager, UI/UX designer, developer) in a development team can be ineffective. For example, it is difficult to completely explore all {the} functionalities of the app in a short period of time.
Inspired by the conception of {storyboard} in movie production, we propose a system, StoryDroid, to automatically generate the storyboard for Android apps, and assist different roles to review apps efficiently. 
Specifically, StoryDroid extracts the activity transition graph and leverages static analysis techniques to render UI pages to visualize the storyboard with the rendered pages. 
The mapping relations between UI pages and the corresponding implementation code (e.g., layout code, activity code, and method hierarchy) are also provided to users. Our comprehensive experiments unveil that StoryDroid is effective and indeed useful to assist app development. 
The outputs of StoryDroid enable several potential applications, such as the recommendation of UI design and layout code.
\end{abstract}
	
	
\section{Introduction}
Mobile apps now have become the most popular way of accessing the Internet as well as performing daily tasks, e.g., reading, shopping, banking and chatting~\cite{web:mobileDesktop}. 
Different from traditional desktop applications, mobile apps are typically developed under the time-to-market pressure and facing fierce competitions --- over 3.8 million Android apps and 2 million iPhone apps are striving to gain users on Google Play and Apple App Store, the two primary mobile app markets~\cite{web:appNumber}.

Therefore, for app developers and companies, it is crucial to perform extensive competitive analysis over existing apps with similar purposes{~\cite{guo2017automated,arbon2014app,web:competitorAnalysis,fox2017mobile}.}
This analysis helps understand the competitors' strengths and weaknesses, and reduces market risks before development.
Specifically, it identifies common app features, design choices, and potential customers.
Moreover, researching similar apps also helps developers gain more insight on the actual implementation, given that delivering commercial apps can be time-consuming and expensive~\cite{web:appCost}.

Typically, to achieve the aforementioned analysis, a freelance developer or a product manager (PM) in a tech company has to download the apps from markets, install them on mobile devices, and use them back-and-forth to identify what she is interested in{~\cite{guo2017automated,arbon2014app,fox2017mobile,web:competitorAnalysis}}.
However, such manual exploration can be painstaking and ineffective. 
For example, if a tech company plans to develop a social media app, 245 similar apps on Google Play will be under review. It is overwhelming to manually analyze them --- register accounts, feed specific inputs if required, and record necessary information (e.g., what are the main features, how are the app pages connected).
Additionally, commercial apps can be too complex to manually uncover all functionalities in a reasonable time~\cite{azim2013targeted}. 
For UI/UX designers, the same exploration problem still remains when they want to get inspiration from similar apps' design.
In addition, the large number of user interface (UI) screens within the app also makes it difficult for designers to understand the relation and navigation between pages.
For developers who want to get inspiration from similar apps, it is difficult to link the UI screens with the corresponding implementation code --- the code can be separated in static layout files as well as a large piece of functional code.

\begin{figure}
		\center
		\includegraphics[width=0.45\textwidth]{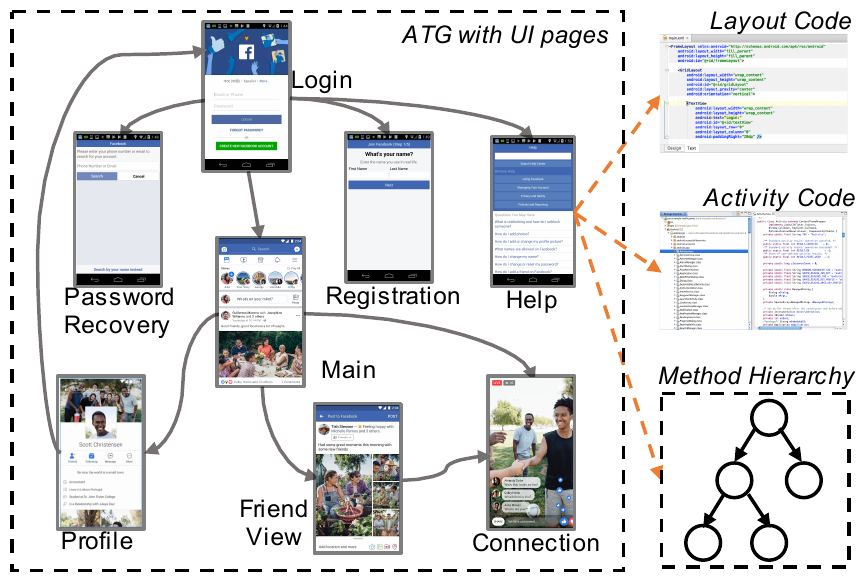}
		\vspace{-1mm}
		\caption{The storyboard diagram of an Android app}
		\label{fig:demo}
\end{figure}

Inspired by the conception of \emph{storyboard}\footnote{``Storyboard'' was developed at Walt Disney Productions, including a sequence of drawings typically with some directions and dialogues, representing the shots planned for a movie or television production.} in movie industry~\cite{finch1995art}, we intend to generate the \emph{storyboard} of an app to visualize its key app behaviors.
Specifically, we use activities (i.e., UI screens) to characterize the ``scenes'' in the storyboard, since activities represent the intuitive impression of the apps in a full-screen window and are the most frequently used components for user interactions~\cite{activity}.
Fig.~\ref{fig:demo} shows the storyboard diagram of \emph{Facebook} (one of most popular social media apps), which includes the activity transition graph (ATG) with UI pages, the detailed layout code (e.g., static and dynamic), the functional code of each activity (\emph{Activity Code}), and method call relations within each activity (\emph{Method Hierarchy}).
Based on this storyboard, PMs can review a number of apps in a short period of time and propose more competitive features in their own app.\footnote{The main target is to help PMs, developers, designers understand and get inspiration from existing apps, instead of directly distributing any part of the code for developing apps for commercial purpose.} 
UI designers can obtain the most related UI pages for reference. 
And developers can directly refer to the related code to improve development efficiency.

However, generating the storyboard is challenging.
First, ATGs are usually incomplete due to the limitation of static analysis tools~\cite{octeau2015composite,azim2013targeted}. 
Second, to identify all UI pages, a pure static approach may miss parts of UIs that are dynamically rendered (see Section~\ref{sec:background}), whereas a pure dynamic approach~\cite{monkey, su2016fsmdroid, su2017guided, chen2018ui, chen2019codegeneration} may not be able to reach all pages in the app, especially those requiring login.
Third, the obfuscated activity names lack the semantics of corresponding functionalities, making the storyboard hard to understand.

To overcome these challenges, we propose a system, \tool, to automatically generate the storyboard of apps in three main phases: 
(1) \emph{Activity transition extraction}, which extracts ATG from the apks, especially the transitions in fragments~\cite{fragment} (components of activities) and inner classes~\cite{inner}, making ATG more complete.
(2) \emph{UI page rendering}, which first extracts the dynamic components (if any) for each UI page and embeds them into the corresponding static layout. It then renders each UI page statically based on the static layout files.
(3) \emph{Semantic name inferring}, which infers the semantic names for the obfuscated activity names by comparing the layout hierarchy with the ones in our database which is constructed from 4,426 open source apps.
Through these analyses, \tool provides a systematic solution for exploring and understanding an app from different points of view.

We evaluate \tool on 100 apps (50 open-source and 50 closed-source apps) from the following two aspects: 
(1) effectiveness evaluation of each phase; 
(2) usefulness evaluation of the visualization outputs.
The experimental results show the effectiveness of \tool in extracting activity transitions, especially in fragments and inner classes. 
\tool extracts nearly 2 times more activity transitions than the state-of-the-art ATG extraction tool (i.e., \textsc{\small IC3}~\cite{octeau2015composite}) on both open-source apps and closed-source apps.
Besides, \tool significantly outperforms the state-of-the-art dynamic testing tool (i.e., \textsc{\small Stoat}~\cite{su2017guided}) on activity coverage for both open-source apps (87\% on average) and closed-source apps (74\% on average).
On average, our rendered images achieve 84\% similarity compared with the ones that are dynamically obtained by \textsc{\small Stoat}. 
And \tool can infer the semantic names at a high accuracy (92\%).
In addition, the user study shows that with the help of \tool, the activity coverage has a significant improvement compared with exploration without \tool, and users can find the layout code for the given UI pages more accurately and efficiently.
Apart from the fundamental usefulness, we also discuss several additional practical applications based on the outputs of \tool, such as the recommendation of UI design and layout code and guiding app regression testing.

In summary, we make the following contributions:

\begin{itemize}[noitemsep,topsep=0pt,leftmargin=*]
\item This is the first research work to automatically generate the storyboard of Android apps. It assists app development teams including PMs, designers, and developers to quickly have a clear overview of other similar apps. 

\item We propose a novel algorithm to extract relatively complete ATG, render UI pages statically, and infer activity names for obfuscated apps. These technical contributions are general for both open-source and close-source Android apps.

\item Our experiments demonstrate not only the accuracy of the generated storyboard, but also the usefulness of our storyboard for assisting app development.

\item This is a fundamental work to enable the construction of a large-scale database of app storyboards, as the overall approach is based on static program analysis.
Such a database can expand the horizon of current mobile app research by enabling lots of future work such as extracting commonalities across apps, recommending UI code, design, and guiding app testing. 

\end{itemize}

\section{Motivating Scenario}
We detail the typical app review process{~\cite{devprocess,typical,guo2017automated,arbon2014app,fox2017mobile}} with our \tool for Android apps in term of different roles in the development team. 
Eve is a PM of an IT company. 
Her team plans to develop an Android social app.
In order to improve the competitiveness of the designed app, she searches hundreds of similar apps (e.g., Facebook, Instagram, Twitter) based on the input keywords (e.g., social, chat) from Google Play Store.
She then inputs all of the URLs of these apps into \tool which automatically download all of these apps with Google Play API~\cite{api}.
\tool further generates the storyboard (e.g., Fig~\ref{fig:demo}) of all these apps and displays them to Eve for an overview.
By observing these storyboards together, she easily understands the storyline of these apps, and spots the common features among these apps such as registering, searching, setting, user profile, posting, etc.
Based on these common features, Eve comes up with some unique features which can distinguish their own app from existing ones.

Alice, as a UI/UX designer, needs to design the UI pages according to Eve's requirements. 
With our \tool, she can easily get not only a clear overview of the UI design style of related apps, but also interaction relations among different screens within the app.
Then, Alice can develop the UI and user interaction of her app inspired by others' apps~\cite{web:designer1, web:designer2}.

Bob is an Android developer who needs to develop the corresponding app based on Alice's UI design. 
Based on Alice's referred UI design in the existing app, he can also refer to that app with the help of our \tool.
By clicking the UI screen of each activity in the storyboard, \tool returns the corresponding UI implementation code no matter it is implemented with pure static code, dynamic code, or hybrid ones.
{To implement their own UI design, he can refer to the implemented code and customize it based on their requirement.
That development process is much faster than starting from scratch.
In addition, Bob may also be interested in certain functionality within a certain app. By using \tool, he easily locates the logic code.}

\section{Preliminaries}\label{sec:background}
In this section, we briefly introduce the concepts of Android UI, layout types of UI and special views for populating data.

\subsection{Android Activity and Fragment}
There are four types of components in Android apps (i.e., \texttt{\small Activity}, \texttt{\small Service}, \texttt{\small Broadcast}, \texttt{\small Receiver}). 
Specifically, \texttt{\small Activity}~\cite{activity} and \texttt{\small Fragment}~\cite{fragment} render the user interface and are the visible parts of Android apps. \texttt{\small Activity} is a fundamental component for drawing the screens which users can interact with. 
\texttt{\small Fragment} represents a portion of UIs in the activities, which contributes their own UI to certain activities.
\texttt{\small Fragment} always depends on an \texttt{\small Activity} and cannot exist independently. 
A \texttt{\small Fragment} can also be reused in multiple activities and an activity may contain multiple fragments based on the screen size, with which we can create multi-panel UIs to adapt to mobile devices with different screen sizes.

\begin{figure}
	\centering
	\includegraphics[width=0.45\textwidth]{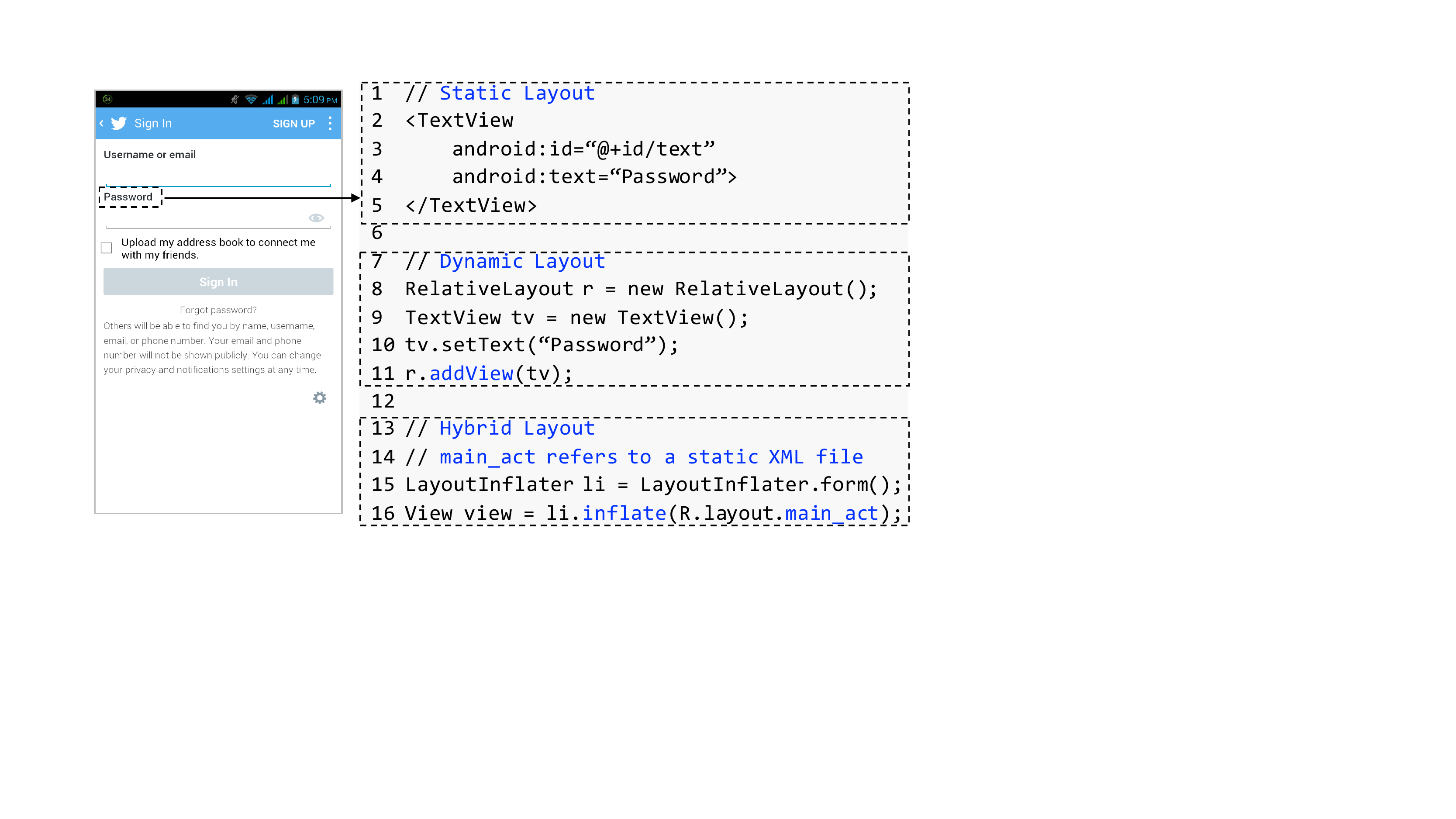}
	\vspace{-1mm}
	\caption{Three types of GUI layout}
	\label{fig:layout}
\end{figure}

\subsection{UI Layout}
A user interface, rendered by \texttt{\small Activity} and \texttt{\small Fragment}, requires a UI layout to draw a window for users. We take a high-level look at three layout types in Android apps. 
Fig.~\ref{fig:layout} shows the login UI page of a famous social app, Twitter~\cite{twitter}, where the component (e.g., \texttt{\small TextView}) can be implemented in three different layout types.
	
\noindent \textbf{Static layout}, which relies on the static layout files (i.e., XML) in the apk.  
The UI pages of the app are rendered by these XML files. \texttt{\small ViewGroup} and \texttt{\small View} are the basic elements of user interfaces in UI layout files. 
\texttt{\small ViewGroup} is a container which holds other \texttt{\small ViewGroups} and \texttt{Views}. 
The GUI code must contain a root node with \texttt{\small ViewGroup} (e.g., \texttt{\small RelativeLayout} and \texttt{\small LinearLayout}). 
When the \texttt{\small ViewGroup} is defined, developers can add additional \texttt{View}s (e.g., \texttt{\small EditText}, \texttt{\small TextView}) as children elements to gradually build a hierarchy that defines the page layout. 
GUI components contain multiply attributes (e.g., id, text, width) as shown in Fig.~\ref{fig:layout} (lines 2-5).
	
\noindent \textbf{Dynamic layout}, which allows developers to instantiate layout components at runtime using Android API (i.e., \texttt{\small addView()}).  
Developers can create components and manipulate their attributes in Java code, e.g., \textsf{\small TextView.setText}(``\textsf{\small Password}'') in Java code is equivalent to \textsf{\small android:text}=``\textsf{\small Password}'' in layout file as shown in Fig.~\ref{fig:layout} (lines 7-11).
	
\noindent \textbf{Hybrid layout}, which defines some default components in static layout files. These layout can be reused dynamically by invoking \texttt{\small LayoutInflater.inflate} as shown in Figure~\ref{fig:layout} (lines 14-15).
Developers can also manipulate the attributes of the defined components. For example, they can modify the content of \texttt{\small TextView} in Java code.

To investigate the proportion of dynamic and hybrid layout used in Android apps, we randomly select 1,000 apps from the top 10,000 Google Play apps with the most number of downloads. 
We use the specific Android APIs (i.e., \texttt{\small addView()} and \texttt{\small inflate()}) to distinguish if the app contains dynamic/hybrid layout types to draw the UI pages.
The result of our study unveils that 62.3\% apps use dynamic/hybrid layout. 
The reason for such frequent usage of dynamic/hybrid layout is that with the help of them, the views are separated from the view model in the XML file, developers can change the layout without recompiling to adapt to the app's runtime state. 
	
\begin{figure}[t]
	\scriptsize
	\centering
	\begin{lstlisting}[language=Java]
	public class PartList extends Activity{
	  ListView lv = (ListView) findViewById(R.id.list);
	  ArrayAdapter adapter = new ArrayAdapter(this,R.layout.list_view,data);
	  lv.setAdapter(adapter); ...}\end{lstlisting}
	\vspace{-3mm}
	\caption{Simplified code snippet of Adapter}
	\label{fig:adapter}
\end{figure}
	
\subsection{Data Population}
\texttt{\small{Adapter}}~\cite{adapter} is a bridge between the \texttt{\small AdapterView} and the underlying data for the view. 
It also provides the layout (e.g., \texttt{\small ListView}, \texttt{\small GridView}, \texttt{\small ViewPager}) with the data, which is usually loaded from a local database or remote server. 
\texttt{\small{Adapter}} enables these UI components (e.g., \texttt{\small ListView}) to provide a list of scrollable items for the selection purpose.
Fig.~\ref{fig:adapter} shows an example of \texttt{\small Adapter}. It instantiates an adapter with a layout (\texttt{\small ListView}) and associates it with \texttt{\small data}. The data is then displayed in a \texttt{\small ListView} via adapter.
The \texttt{\small AdapterView} is part of the user interface layout and should be extracted from the source code for static rendering.

\begin{figure}
	\centering
	\includegraphics[width=0.425\textwidth]{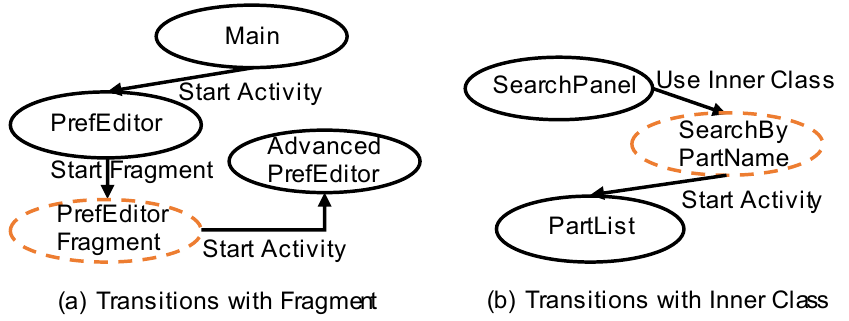}
	\vspace{-1mm}
	\caption{Transitions between activities and fragment, inner class}
	\label{fig:atg}
\end{figure}

\section{Our Approach}
\begin{figure*}
	\centering
	\includegraphics[width=0.9\textwidth]{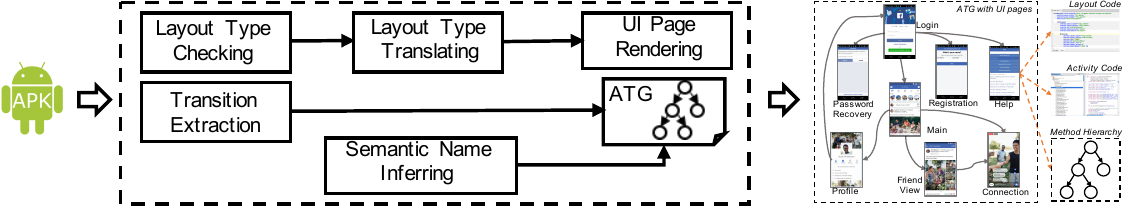}
	\vspace{-1mm}
	\caption{Architecture of \tool}
	\label{fig:workflow}
	\vspace{-6mm}
\end{figure*}

\begin{figure}[t]
	\scriptsize
	\centering
	\vspace{-2mm}
	\begin{lstlisting}[language=Java]
	public class PrefEditor{... //Using replace/add
	  PrefEditorFragment pref = new PrefEditorFragment();
	  FragmentTransaction.replace(R.id.content,pref);
	  FragmentTransaction.commit(); }
	public class PrefEditor{... // Using setAdapter
	  ViewPager.setAdapter(getSupFragmentManager(),
	  new PrefEditorFragment()); }\end{lstlisting}
	\vspace{-3mm}
	\caption{Simplified code snippet of Fragment}
	\label{fig:fragment}
	\vspace{-3mm}
\end{figure}

\begin{figure}[t]
	\scriptsize
	\centering
	\vspace{-1mm}
	\begin{lstlisting}[language=Java]
	public class SearchPanel{...
	private class SearchByPartName extends Asynctack<>{...
	  Intent intent = new Intent
	(MainActivity.this,PartList.class);
	  startActivity(intent); } }\end{lstlisting}
	\vspace{-3mm}
	\caption{Simplified code snippet of Inner Class}
	\label{fig:inner}
\end{figure}

\tool takes an \emph{apk} as input and outputs the visualized storyboard (${S}$) for the app. 
Fig.~\ref{fig:workflow} shows the three main phases of \tool: (1) \emph{Transition Extraction}, which enhances the ATG extraction ability of \textsc{\small IC3}~\cite{octeau2015composite}, especially for fragments and inner classes. \textsc{\small StoryDroid} leverages control- and data-flow analysis to obtain relatively complete ATG. (2) \emph{UI Page Rendering}, which translates dynamic and hybrid layout to static layout (if needed) to render UI pages that users interact with. 
(3) \emph{Semantic Name Inferring}, which infers the semantic name for the obfuscated activity names by layout comparison.

\SetKwInput{KwInput}{Input}
\SetKwInput{KwInput}{Input}
\SetKw{Let}{let}
\SetKw{Continue}{continue}

\begin{algorithm2e}[t]
	\footnotesize
	\setcounter{AlgoLine}{0}
	\caption{ATG and Rendering Source Extraction}
	\label{algo:static}
	\DontPrintSemicolon
	\SetCommentSty{mycommfont}
	\KwIn{$apk$}
	\KwOut{$atg$: Activity transition graph, $adapters$: Adapter mappings, $layout\_type$: i.e., static, dynamic or hybrid}
	
	$atg$ $\leftarrow \emptyset$, $adapters$ $\leftarrow \emptyset$  \;
	$cg$ $\leftarrow$ getCallGraph($apk$) \;
	$layout\_type$ $\gets$ getLayoutType($apk$)\;
	$all\_classes$ $\gets$ getAllClasses($apk$) \;
	\ForEach{$c$ $\in$ $all\_classes$}{
		
		$methods$ $\gets$ getClassMethods($c$) \;
		\ForEach{$m$ $\in$ $methods$}{
			
			\If{hasActivityTransition($m$)}{
				$callee\_act$ $\gets$ getTargetAct($m$)\;
				
				\If{isInnerClass($c$)}{
					$caller\_act$ $\gets$ outerClass($c$) \;
					$atg$.addPair($caller\_act$, $callee\_act$)\;
				}
				
				\ElseIf{isInFragment($m$)}{
					$atg$.addPair($caller\_frag$, $callee\_act$)\;
				}
				\Else{
					$caller\_acts$ $\gets$  getCallerAct($m$, $cg$)\;
					\ForEach{$act$ $\in$ $caller\_act$}{
						$atg$.addPair($act$, $callee\_act$)\;
					}
				}	
			}
			\tcp*[h]{\scriptsize \texttt{get rendering sources and optimize $atg$}}\;
			\If{startFragment($m$)}{
				$caller\_acts$ $\gets$  getCallerAct($caller\_frag$)\;
				\ForEach{$act$ $\in$ $caller\_acts$}{
					$atg$.addPair($act$, $callee\_frag$)\;
				}
				updateATGIfNeeded($atg$)
			}	
			
			\tcp*[h]{\scriptsize \texttt{get rendering sources}}\;
			\If{hasAdapter($m$)}{
				\tcp*[h]{\scriptsize \texttt{e.g., ListView, RecyclerView, ViewPager}}\;
				$view\_type$ $\gets$ getViewType($m$)\;
				$layout\_file$ $\gets$ BackwardAnalysis($m$)\;
				$adapters$.add($c$, $view\_type$, $layout\_file$)\;
			}
		}
	}
	\Return $atg$, $adapters$, $layout\_type$
	
\end{algorithm2e}

\subsection{Transition Extraction}
Before extracting activity transitions in inner classes and fragments, we illustrate the transitions in them.
Fig.~\ref{fig:atg} (a) is the sub ATG of \texttt{\small Vespucci}~\cite{vespucci}, a map editor.
Firstly, activity \textsf{\small  Main} starts \textsf{\small PrefEditor}, in which \textsf{\small PrefEditorFragment} is started. And \textsf{\small PrefEditorFragment} further starts \textsf{\small AdvancedPrefEditor}.
Specifically, as shown in Fig.~\ref{fig:fragment}, fragments can be added to an activity in two ways: 
(1) by invoking fragment modification API calls, e.g., ``\texttt{\small replace()},'' ``\texttt{\small add()},'' and further leveraging ``\texttt{\small FragmentTransation.commit()}'' (lines 3-4) to start the fragment; 
(2) By using ``\texttt{\small setAdapter}'' (line 6) to display the fragment in a certain view (e.g., \texttt{\small ViewPager}). The started \textsf{\small PrefEditorFragment} then starts a new activity (i.e., \textsf{\small AdvancedPrefEditor}).
Fig.~\ref{fig:atg} (b) shows the sub ATG of \texttt{\small ADSdroid}, where \textsf{\small SearchPanel} uses an inner class \textsf{\small SearchByPartName} to handle time-consuming operations as shown in Fig.~\ref{fig:inner}.
After finishing the task, it starts an activity \textsf{\small PartList} by invoking ``\texttt{\small StartActivity()}'' (line 5). 
In this example, our goal is to extract activity transitions: \textsf{\small Main}$\rightarrow$\textsf{\small PreEditor}, \textsf{\small PreEditor}$\rightarrow$\textsf{\small AdvancedPreEditor}, and \textsf{\small SearchPanel}$\rightarrow$\textsf{\small PartList}. 

Algorithm~\ref{algo:static} details the extraction of ATG and resources for rendering, including the layout type (i.e., static, dynamic, or hybrid) and the adapter mapping relations. Specifically, it takes as input an ${apk}$, and outputs the activity transition graph (${atg}$), adapter mappings (${adapters}$), and layout type (${layout\_type}$). 
We first initialize $atg$ as an empty set (line 1), which stores the activity transitions gradually. We then generate the call graph ($cg$) of the given apk, obtain the layout type by analyzing the existence of dynamic layout loading APIs (lines 2-3). For each method ($m$) in each class ($c$), if there exists an activity transition, we first get the target activity ($callee\_act$) by analyzing the data in \texttt{\small Intent} (lines 5-9).
If the method ($m$) is in an inner class, we regard the outer class as the activity that starts the target activity and add the transition to $atg$ (lines 10-12).
Take Fig.~\ref{fig:atg}~(b) as an example,
we add an edge \textsf{\small SearchPanel}$\rightarrow$\textsf{\small PartList} to $atg$. 
If $m$ is in a fragment, we construct the relation between the fragment ($caller\_frag$) and the target activity (lines 13-14). Note that this relation does not represent the actual activity transition, we optimize it by identifying the activities that start the fragment in lines 19-22. This relation is used for both ATG construction and UI page rendering. After that we update $atg$ by merging fragment relations to construct the actual activity transitions (line 23). 
For example, in Fig.~\ref{fig:atg}~(a), we first obtain the relations \textsf{\small PrefEditorFragment}$\rightarrow$\textsf{\small AdvancedPreEditor}, \textsf{\small PrefEditor}$\rightarrow$\textsf{\small PrefEditorFragment}, then we merge it to \textsf{\small PrefEditor}$\rightarrow$\textsf{\small AdvancedPreEditor} to represent the actual activity transition.
For method $m$ that is neither in an inner class nor a fragment, we backward traverse $cg$ starting from $m$ to obtain all the activities that start the target activity ($callee\_act$), then add them to $atg$ (lines 15-18).

In addition, to complement the UI layout that need to load data from data providers (e.g., \texttt{\small ContentProvider}, \texttt{\small Preference}) using different types of views (e.g., \texttt{\small ListView}, \texttt{\small RecyclerView}, \texttt{\small ViewPager}), 
we first identify the method that uses \texttt{\small Adapter} and obtain the correponding view type ($view\_type$) (lines 24-25). 
We then utilize backward data-flow analysis on the adapters to track the corresponding layout files, which pinpoints the layout/activity that will be embedded with data displayed in $view\_type$. 
We define each mapping relation as a tuple $\langle $activity$, $view\_type$, $layout$ \rangle $ and save them in $adatpers$ for UI rendering.
Take Fig.~\ref{fig:adapter} as an example, we denote the relation as $\langle$PartList$, $ListView$, $list\_view$\rangle $.

\subsection{UI Page Rendering}

We propose to statically render the UI pages due to the limitation of data dependence between different activities when using dynamic tools for UI page rendering (e.g., require login or special input to reach another activity). The activities we rendered are the initial state of each activity.
As for apps that only use static layout to display UI pages, we can directly extract the corresponding layout file for rendering. 
However, as for dynamic/hybrid layout, we need to resolve two challenges: (1) converting dynamic/hybrid layout to static layout since we cannot render the corresponding UI pages accurately with incomplete layout; (2) filling the dynamic data loading area with dummy data (i.e, text, image) since we cannot render the corresponding components that load dynamic data from remote server which involves backend code. To tackle these problems, we first statically analyze the activity source code and identify the logic that is relevant to layout population, including adding new views and modifying the parameters of views. We then convert the dynamic layouts in the target app to static layouts. 
Moreover, as our approach is based on static analysis, we cannot obtain the real image which is loaded dynamically such as \texttt{\small ListView} and \texttt{\small GridView}. Instead of keeping that position plain, we fill in that position with dummy images so that users can directly discriminate it from the plain background. Otherwise, the image position may be preempted by other components of the same page.
We associate the dummy data with the identified adapters in Algorithm~\ref{algo:static} to display data in different styles.
With the translated static layouts and dummy dynamic data, we compile them into an apk to render UI pages.

\begin{algorithm2e}[t]
	\footnotesize
	\setcounter{AlgoLine}{0}
	\caption{UI Page Rendering}
	\label{algo:render}
	\DontPrintSemicolon
	\SetCommentSty{mycommfont}
	\KwIn{$atg^*$: the transitions that contain activities and fragments, $adapters$, $layout\_type$}
	\KwOut{$pages$: UI pages}
	
	$act\_frag \gets$ getActivityAndFragment($atg^*$) \;
	\ForEach{$act \in act\_frag$}{
		$layout$ $\gets$ copy($act.xml$) \;
		
		\If{$layout\_type$ $\neq$ ``static''} {
			$par\_layout \gets$ getParentLayout($act$) \;
			$methods \gets$ getAllMethods($act$) \;
			\ForEach{$m \in methods$}{
				\If{$m$ $\neq$ ``onCreate''}{
					{continue} \;
				}
				$compt, attr \gets$ DataFlowAnalysis($m$) \;
				$par\_layout$.add($compt$) \;
				\tcp*[h]{\scriptsize \texttt{add corresponding attributes}}\;
				\If{isMethod($attr$)}{
					$compt$.add(DataFlowAnalysis ($attr$)) \;
					
				} \Else{
					$element$ $\gets$ getElementByName($attr$)\;
					$compt$.add($element$)\;
				}
				
				$break$\;
				
			}
			$act\_layout$.add($par\_layout$) \;
			saveLayout($act\_layout$)\;
		}
		
	}
	
	\ForEach{$adapter$ $\in$ $adapters$}{
		$activity$ $
		\gets$ $match\_activity(adapter)$
		
		$modify\_layoutFile(activity, adapter.type)$
	}
	
	$apk$ $\gets$ bulidApk($act\_frag$) \;
	$pages$ $\gets$ getScreenshot($apk$) \;
	\Return $pages$
\end{algorithm2e}

Algorithm~\ref{algo:render} details the UI page rendering process. It takes as input the activity transition graph (${atg}^*$), adapter mapping relations (${adapter}$), layout type (${layout\_type}$), and outputs the rendered UI pages. 
We first extract all the activities and fragments from $atg^*$ since we need to render both activity UIs and fragment UIs to make it closer to the real UIs.
For each activity/fragment ($act$), if $act$ uses the static layout, we directly make a copy of the corresponding layout file for further rendering or modification (line 3). 
However, if the app uses dynamic/hybrid layout, we aim to add the dynamic components together with their attributions to the corresponding parent layout. Specifically,
we first extract and pinpoint the parent layout (e.g., \texttt{\small LinearLayout}) (line 8). We then traverse each method ($m$) to identify the dynamic component $compt$ (e.g., \texttt{\small EditText}, \texttt{\small TextView}) and the corresponding attributes $attr$ (e.g., text, height, width) by keywords matching (i.e., \texttt{\small inflate} and \texttt{\small addView}) and forward analyzing the data flow of view-related variables in \texttt{\small onCreate()}. We then add $compt$ to the corresponding parent layout (lines 7-11). As for the attribute, if it is obtained by invoking another method, we leverage data-flow analysis to reach the definition of the attribution, and attach it to the corresponding component (lines 12-13). Otherwise, we identify the corresponding element in the resource files by the attribute name via \texttt{\small getElementByName()}, and also attach it to $compt$ (lines 15-16). 
After the modification to the parent layout of the dynamic components, we attach the new $par\_layout$ to $act\_layout$ and save it as layout format for further rendering (lines 18-19).
Take Fig.~\ref{fig:layout} as an example, our method is able to translate the dynamic/hybrid layouts to static layouts (i.e., XML format).

In addition, as for data display using adapters, we match each adapter to the activity and modify the corresponding layout by embedding the view types, e.g., \texttt{\small ListView}, \texttt{\small RecyclerView} (lines 20-22) as well as dummy data. We finally get a view tree and ensure the corresponding attributes are added from the source code.
At last, we build an apk to render all the activities and fragments, and take screenshots for each UI (lines 23-24) in a real app.

\subsection{Semantic Name Inferring}\label{sec:label}
In Android app development, activity names of apps are recommended to contain semantic meanings and end with ``\emph{Activity}''~\cite{naming}, thus we assume that the defined activity names by developers have basic semantic meanings of the corresponding functionality. 
To verify this assumption, we randomly download 1,000 apps from F-Droid~\cite{fdroid}, extract all the activity names from each app, and finally get 6,767 activity names.  
We manually investigate the activity names and observe that most of activity names have semantic meanings.
However, Android obfuscation techniques are often used in Google Play apps to protect their security~\cite{dong2018understanding}. 
The activity names will be translated to simple words like ``a,'' ``b,'' and ``c,'' totally losing their practical semantic meanings. 
Those obfuscated names significantly hinder users' understanding of our storyboard.
To tackle this problem, we propose to automatically infer the semantic names for obfuscated activities by comparing the layout hierarchy (i.e., \texttt{\small ViewGroup} and \texttt{\small View}) with those of the existing activities since activity layout files are not obfuscated in apps. 
In this paper, we consider the activity names whose length is less than three letters as obfuscated ones.
To infer the semantic names for obfuscated activities, we aim to learn from the activity names of open source apps.

\begin{figure}
	\centering
	\begin{subfigure}[t]{0.28\textwidth}
		\centering
		\includegraphics[width=4.35cm]{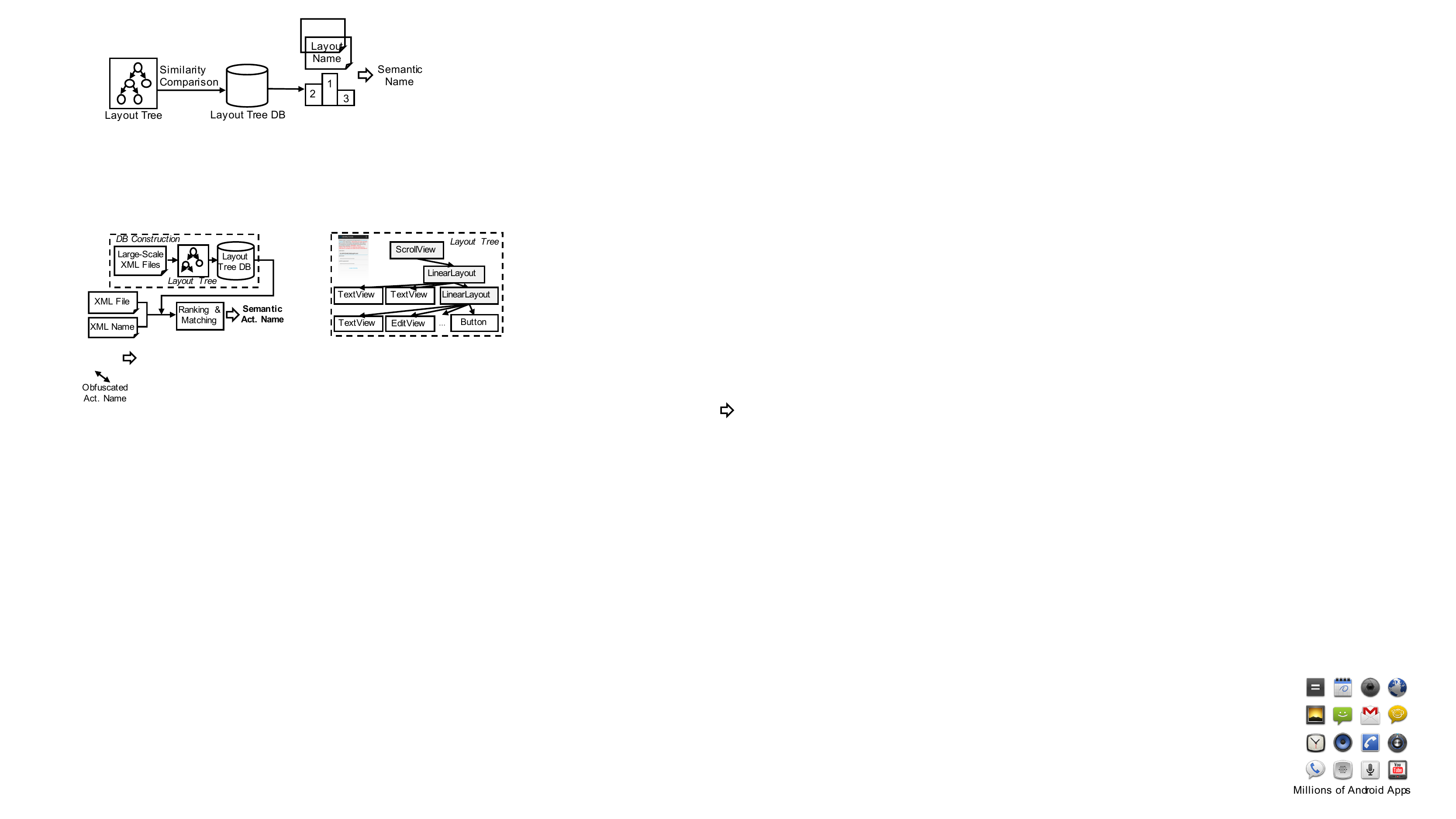}
		\caption{Semantic name inferring}
		\label{fig:label}
	\end{subfigure}%
	\begin{subfigure}[t]{0.21\textwidth}
		\centering
		\includegraphics[width=3.7cm]{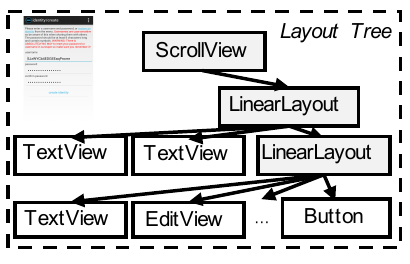}
		\caption{An example of layout tree}
		\label{fig:tree}
	\end{subfigure}
	\vspace{-1mm}
	\caption{Workflow of semantic name inferring}
	\label{fig:infer}
\end{figure}

To achieve this, as shown in Fig.~\ref{fig:infer} (a), we crawl all the apps on F-Droid (4,426 in total) and build a large database based on the layout hierarchies of the apps for similarity comparing. A layout hierarchy is a layout tree (i.e., Fig.\ref{fig:infer} (b)) based on the XML layout code. The root node of the tree is a type of \texttt{\small ViewGroup}, and the other \texttt{\small ViewGroup}s and types of \texttt{\small View}s are the child nodes added to the root node. 
We finally get 13,792 activities with their layout hierarchies. 
Given the XML file (layout file) and XML name of the obfuscated activity, we infer its semantic name by the following three steps: 
(1) \emph{layout hierarchy extraction}, which extracts the layout tree ($T$) from the XML file;
(2) \emph{similarity comparison}, which computes the similarity between $T$ and the trees in our database by leveraging the tree edit distance (TED) algorithm~\cite{zhang1989simple}, which is defined as the minimum-cost sequence of node edit operations that transform one tree into another.
According to a pilot study of 100 randomly selected tree pairs, we define a threshold in TED as 5 and filter the activity names whose TED are less than 5 as candidates.
(3) \emph{ranking and matching}, which infers the activity name from the candidates based on the frequency of each activity name and the corresponding layout names (XML file names). Specifically, we rank the activity names based on their frequency, split the camel-case layout name into multiple single words with the regular expression~\cite{dit2011can}, filter out the general words such as ``activity,'' ``layout,'' and compare it with those in the database by keyword matching. The most matched activity name will be used to rename the obfuscated activity name. However, if the layout name is not matched with any names of the candidates, we rename it with the top frequent name.  
Note that although the layout names are not obfuscated, it is ineffective if we only use them to infer the semantic names since activities using dynamic layouts have no static layouts, thus have no layout names.

\section{Implementation}
We implement \tool as an automated tool, which is written in 3K lines of Java code, and 2K lines of Python code.  \tool is built on top of several off-the-shelf tools: \textsc{\small IC3}, \textsc{\small jadx}~\cite{jadx}, and \textsc{\small Soot}~\cite{soot}. We use \textsc{\small Soot} to extract inputs of UI page rendering, and get the call graphs from apks. Activity transition extraction is built on \textsc{\small IC3} to obtain a comparatively complete ATG. \textsc{\small jadx} is used to decompile the apk to the source code for Android apps. 
We also use \textsc{jadx} to extract XML layout code from each apk, which will not be affected by obfuscation, thus will not affect the semantic name inferring method.
We use data-driven document (D3)~\cite{d3} to visualize \tool's results, which provides a visualized technique based on data in \texttt{\small HTML}, \texttt{\small JavaScript}, and \texttt{\small CSS}. The visualization~\cite{storydroid} contains 4 parts: (1) ATG with activity names and corresponding UI pages; (2) The layout code of each UI page; (3) The functional code of each activity; and (4) the method call relations within each activity.

\section{Effectiveness Evaluation}
In this section, we evaluate the effectiveness of \tool based on the following three research questions: 

\noindent {\bf RQ1:} Can \tool extract a more complete ATG for an app, and achieve better activity coverage than the dynamic testing tool (i.e., \textsc{Stoat})? 

\noindent {\bf RQ2:} Can \tool render UI pages with high similarity compared with the real screenshots?

\noindent {\bf RQ3:} Can \tool infer accurate semantic names for obfuscated activities? 

\subsection{Experimental Setup}
To investigate the capability of handling fragments and inner classes, we self-developed 10 apps\footnote{The 10 apps are available on \url{https://sites.google.com/view/storydroid/}} as our ground-truth benchmark, which cover different features (i.e., activity, fragment, and inner class) we claim to resolve. 
This is the only way to exactly know all the transitions even in fragments or inner classes, and this way is widely used in the literature~\cite{li2014know}. 
We follow the features below to generate the apps: 
(1) apps with transitions only in activities; apps with transitions only in inner classes; apps with transitions only in fragments; apps with transitions in both activities and inner classes; apps with transitions in both activities and fragments; 
(2) we developed 2 apps with 7-10 activities under each rule.
We apply both \textsc{\small IC3} and \tool on the 10 apps and extract the transition pairs respectively. 
Moreover, since the stakeholders are more concerned about popular competitive apps which have already dominated the market. To mimic the real scenario, we randomly download 50 apps from GooglePlay Store with 10M+ installations to demonstrate the effectiveness of \tool on real-world apps.
We compare the number of transition pairs identified by \textsc{\small IC3} and \tool.
We also use these 100 apps to evaluate the activity coverage of \tool and the state-of-the-art dynamic testing tool, \textsc{Stoat}~\cite{su2017guided}, which has been demonstrated to be more effective on app exploration than other tools such as \textsc{\small Monkey}~\cite{monkey} and \textsc{\small Sapienz}~\cite{sapienz}. 
Specifically, we collect all the activities defined in each app from \texttt{\small AndroidManifest.xml}, and compare the number of the rendered activities by \tool and the explored activities by \textsc{Stoat}. 

\begin{table}[t]
	\caption{Capability of handling fragments and inner classes. ({``-'': \textsc{\small IC3} cannot handle})}
	\vspace{-2mm}
	\label{tbl:1}
	\center
	\scriptsize
	\begin{tabular}{c |c| c c c}
		\toprule
		\multicolumn{1}{c}{\bf App ID} & \multicolumn{1}{c}{\bf Feature} & \multicolumn{1}{c}{\begin{tabular}[c]{@{}c@{}}\#{\bf Transition} \\ {\bf Pairs}\end{tabular}} & \multicolumn{1}{c}{\begin{tabular}[c]{@{}c@{}}\#{\bf Identified}\\ {\bf by IC3}\end{tabular}} & \multicolumn{1}{c}{\begin{tabular}[c]{@{}c@{}}\#{\bf Identified}\\ {\bf by StoryDoid}\end{tabular}} \\
		\toprule
		\hline
		1 & \multirow{2}{*}{\bf Activity} & 14 & 14 & 14 \\ \cline{1-1} \cline{3-5} 
		2 &  & 13 & 13 & 13 \\ \hline
		3 & \multirow{2}{*}{\bf Inner Class} & 13 & - & 13 \\ \cline{1-1} \cline{3-5} 
		4 &  & 13 & - & 13 \\ \hline
		5 & \multirow{2}{*}{\bf Fragment} & 13 & - & 13 \\ \cline{1-1} \cline{3-5} 
		6 &  & 13 & - & 13 \\ \hline
		7 & \multirow{2}{*}{\begin{tabular}[c]{@{}c@{}}{\bf Activity}\\ {\bf Inner Class}\end{tabular}} & 13 & 1 & 13 \\ \cline{1-1} \cline{3-5} 
		8 &  & 13 & 1 & 13 \\ \hline
		9 & \multirow{2}{*}{\begin{tabular}[c]{@{}c@{}}{\bf Activity}\\ {\bf Fragment}\end{tabular}} & 10 & 1 & 10 \\ \cline{1-1} \cline{3-5} 
		10 &  & 10 & 1 & 10 \\ \hline
	\end{tabular}
\end{table}

For RQ2, we evaluate the similarity of our statically rendered UI pages with the real UI which are dynamically rendered UI by \textsc{\small Stoat} based on the 100 apps~\cite{chen2018ui}. 
\textsc{\small Stoat} is configured with default settings and given 30 minutes to test each app in order to collect the explored activities. 
We further apply \textsc{\small Stoat} on each app to collect the screenshots of each activity. Since \textsc{\small Stoat} may only explore part of the activities of an app within the given time, we only compare the similarity of the explored activities with the corresponding rendered activities by \tool for a fair comparison. 

For RQ3, to demonstrate the accuracy of inferring semantic names, we randomly select 100 activity names with semantic meanings from the extracted 6,767 activity names in Section~\ref{sec:label} as the ground truth. 
We collect the corresponding layout files from source files, and further utilize our method based on TED~\cite{zhang1989simple} to obtain the semantic names for the 100 activities based on our collection of layout trees (i.e., 13,792 layout files). 
We then compare the results with the original activity names to evaluate the accuracy of our approach i.e., the proportion of semantic names that are correctly inferred. 

\vspace{-1mm}
\subsection{Experimental Results}
\subsubsection{RQ1}
\begin{figure}[t!]
	\centering
	\begin{subfigure}[t]{0.24\textwidth}
		\label{fig:comp1}
		\centering
		\includegraphics[width=3.7cm]{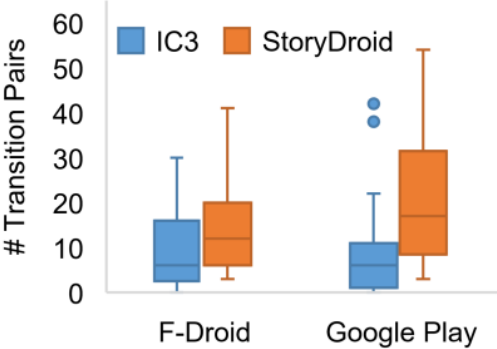}
		\caption{Comparison with \textsc{IC3}}
	\end{subfigure}%
	\begin{subfigure}[t]{0.24\textwidth}
		\label{fig:comp2}
		\centering
		\includegraphics[width=3.7cm]{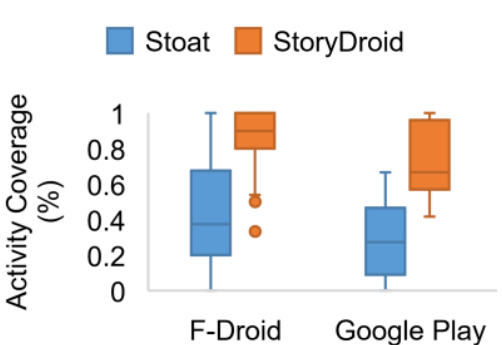}
		\caption{Comparison with \textsc{Stoat}}
	\end{subfigure}
	\vspace{-1mm}
	\caption{Comparison of transitions pairs and activity coverage }
	\label{fig:comp}
\end{figure}

Table~\ref{tbl:1} shows the results of the capability evaluation of handling fragments and inner classes on the 10 ground-truth benchmark apps. 
We can see that \textsc{\small IC3} is able to extract transitions in activities, however, is weak in fragments and inner classes. 
In contrast, \tool can extract transitions with respect to all these features. 
Since we extract transitions by using particular APIs (i.e., \texttt{\small StartActivity}, \texttt{\small StartActivityForReulst}, \texttt{\small StrartActivityIfNeeded}) that start new activities with data-flow analysis, the extracted transitions are more accurate.
The results demonstrate the effectiveness of \tool on extracting activity transitions, especially in fragments and inner classes. 
Moreover, we also evaluate the effectiveness of \tool on real apps compared with \textsc{\small IC3} in Fig~\ref{fig:comp} (a). 
It shows that \tool extracts nearly 2 times more activity transitions and is more effective than \textsc{\small IC3} for both open-source apps and closed-source apps. 

Besides the limitations of inner classes and fragments, transitions in Android system event callbacks are lost in \textsc{\small IC3} according to our observation.
For example, 
(1) \texttt{\small PodListen}~\cite{podlisten} is a podcast player, and there exists an activity transition in the callback method (\texttt{\small DownloadReceiver.onReceive()}), which is called when the \texttt{\small BroadcastReceiver} is receiving an \texttt{\small Intent} broadcast. 
However, \textsc{\small IC3} fails to extract the activity transition. 
(2) \texttt{\small CSipSimple}~\cite{csipsimple} is an online voice communication app, in which the system callback method (\texttt{\small SipService.adjustVolume()}) starts a new activity, but \textsc{IC3} fails to identify the transition. 

Fig.~\ref{fig:comp} (b) depicts the activity coverage results. 
On average, \tool outperforms \textsc{\small Stoat} in terms of activity coverage, achieving 87\% and 74\% coverage on open-source apps and closed-source apps, respectively. 
In addition, \tool costs much less time (i.e., within 3 minutes on average) to extract and render the activities than \textsc{\small Stoat} (i.e., 30 minutes).
\tool does not cover all the activities for some apps due to the following reasons:
(1) the limitation of reverse engineering techniques, some classes and methods cannot be decompiled from the apks, causing failure in the extraction of activity transition and coverage. 
That situation is more severe in closed source apps due to packing~\cite{packers} and obfuscation. 
In our evaluated apps, \texttt{\small PodListen} is the only project in which activity coverage of \tool is lower than that of \textsc{\small Stoat} due to decompiling failure of \texttt{\small SubscribeDialog} fragment.
(2) Another reason is the dead activities (no transitions), such as unused legacy code and testing code in apps.

We further investigate the reasons why \textsc{\small Stoat} achieves low activity coverage: 
(1) Login requirement. For example, \textsc{\small Stoat} fails to explore \texttt{\small Santander} which is a banking app requiring login using password or fingerprint.
(2) Lack of specific events. For example, \texttt{\small Open Training} is a fitness-training app, which can create fitness plans by swiping across the screen. However, \textsc{\small Stoat} does not support such events, resulting low coverage.

\begin{figure}
	\centering
	\begin{subfigure}[t]{0.12\textwidth}
		\centering
		\includegraphics[width=2cm, height=3.15cm]{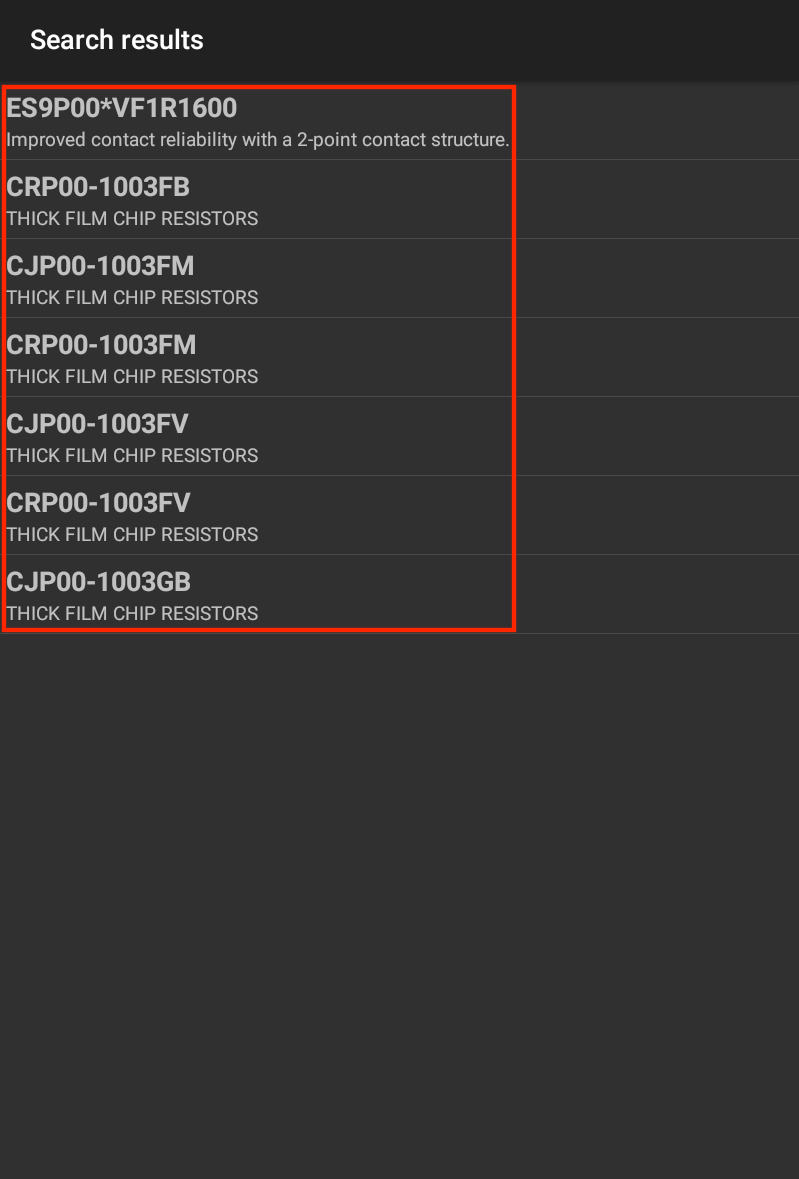}
		\caption{Real page}
	\end{subfigure}%
	\begin{subfigure}[t]{0.12\textwidth}
		\centering
		\includegraphics[width=2cm, height=3.15cm]{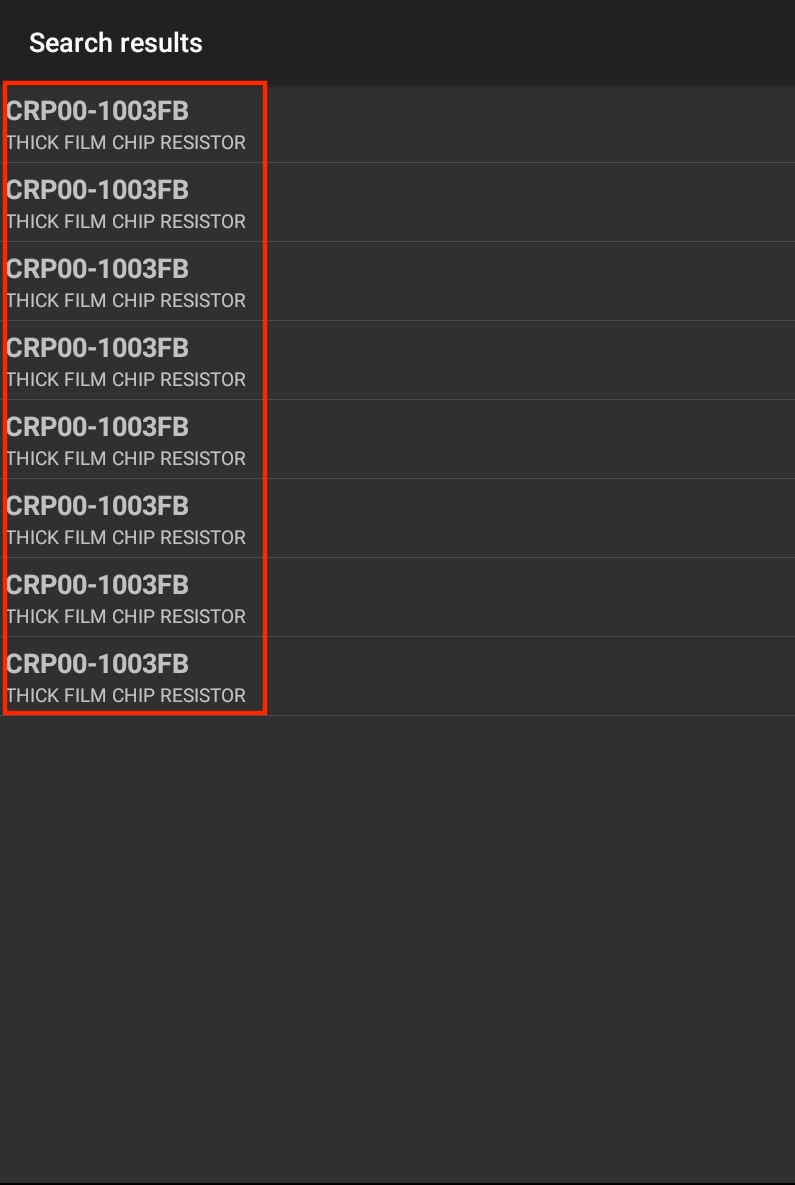}
		\caption{Our rendered}
	\end{subfigure}
	\begin{subfigure}[t]{0.12\textwidth}
		\centering
		\includegraphics[width=2cm]{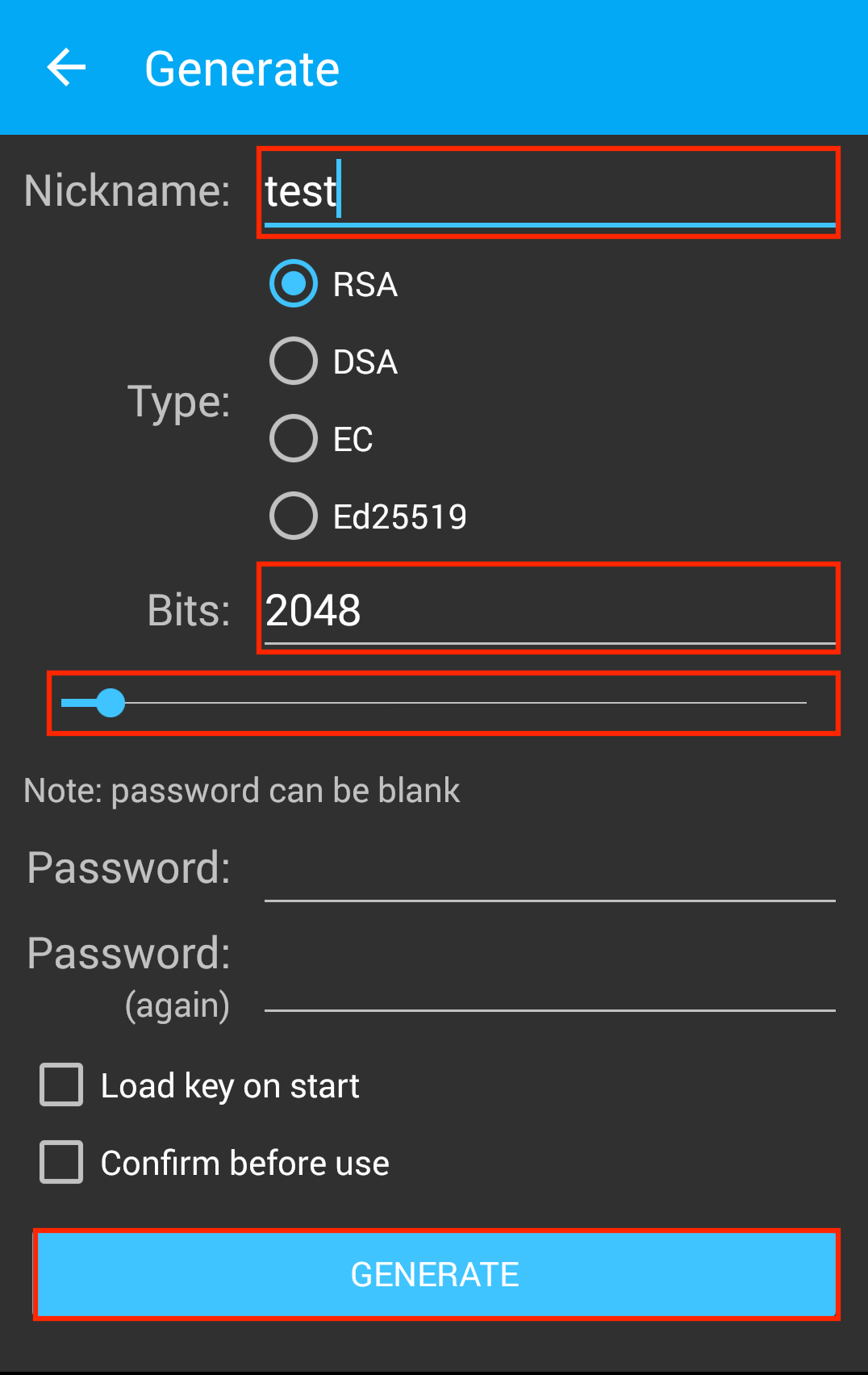}
		\caption{Real page}
	\end{subfigure}%
	\begin{subfigure}[t]{0.12\textwidth}
		\centering
		\includegraphics[width=2cm]{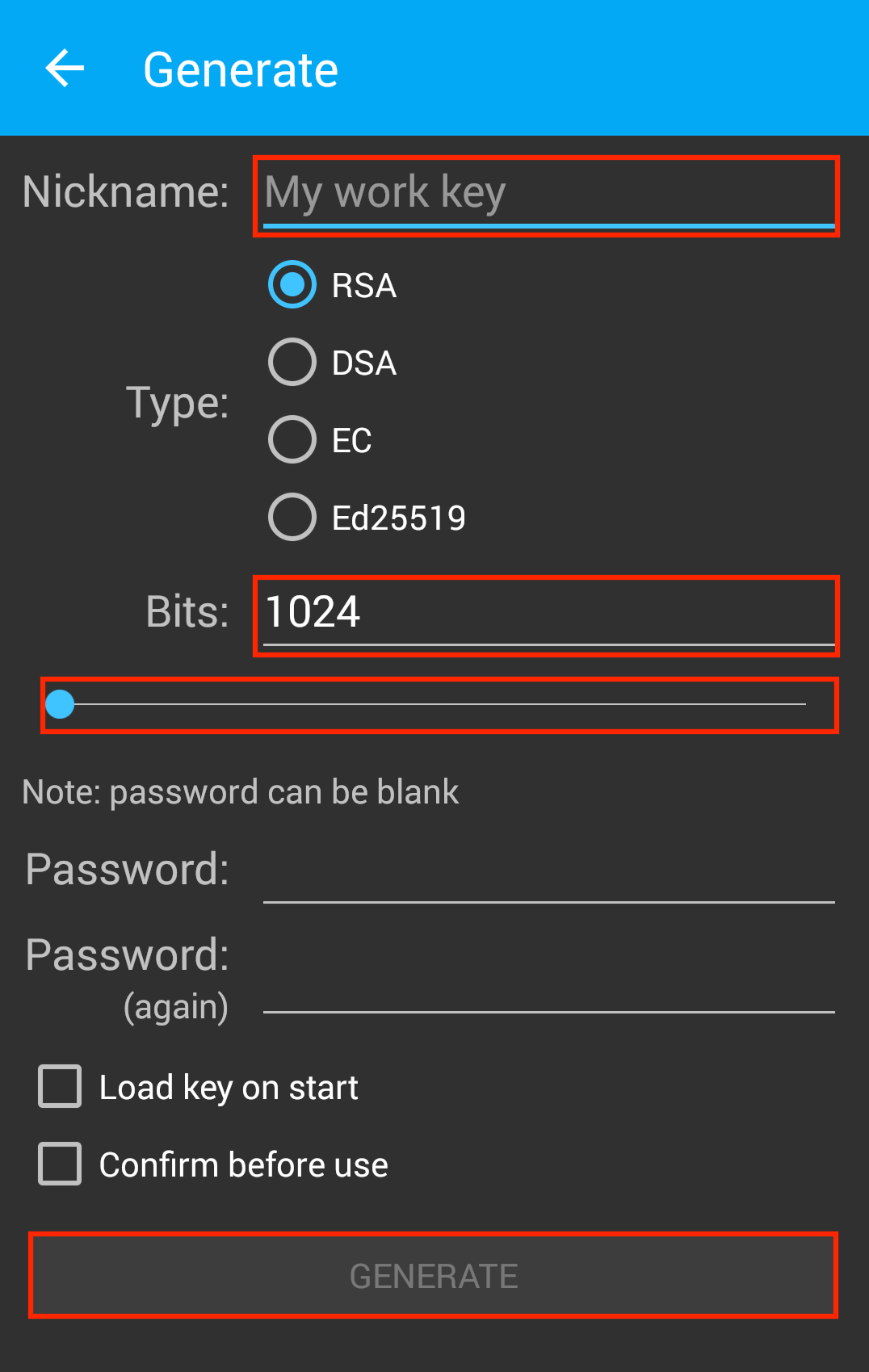}
		\caption{Our rendered}
	\end{subfigure}%
	\vspace{-1mm}
	\caption{Comparison of real pages and rendered pages}
	\label{fig:render}
	\vspace{-2mm}
\end{figure}

\begin{table}
\caption{ Partial results of inferring semantic name for obfuscated activities}
\vspace{-2mm}
\label{tbl:labeled}
\centering
\scriptsize
\begin{tabular}{c c c c}
\toprule
\multicolumn{1}{c}{\begin{tabular}[c]{@{}c@{}}{\bf GroundTruth} \\ {\bf Act. Name}\end{tabular}} & \multicolumn{1}{c}{\begin{tabular}[c]{@{}c@{}}{\bf Rank in} \\ {\bf Candidates}\end{tabular}} & \multicolumn{1}{c}{\begin{tabular}[c]{@{}c@{}}{\bf Corresponding} \\ {\bf XML Name}\end{tabular}} & \multicolumn{1}{c}{\begin{tabular}[c]{@{}c@{}}{\bf Inferred by} \\ \textsc{\scriptsize storydroid}\end{tabular}} \\ \bottomrule
AboutAct. & 1 & about & AboutAct. \\
HelpAct. & 2 & activity\_help & HelpAct. \\ 
\rowcolor{gray!25}PersonalInfoAct. & 3 & content\_extended\_title & \textbf{WizardAct.}\\ 
LoginAct. &  3 & login & LoginAct. \\ 
ContactListAct. & 1 & contact\_list & ContactListAct. \\ 
\rowcolor{gray!25}SearchAct. & 4 & grid\_base & \textbf{Searcher} \\ 
SettingAct. &  1  & setting\_container & SettingAct. \\ 
ShareAct. &  1 & activity\_share & ShareAct. \\
SplashAct. &  3 & activity\_splash & SplashAct. \\
\rowcolor{gray!25}TrackListAct. & 1  & list\_view & {\bf TrackListAct.} \\
\bottomrule
\end{tabular}
\end{table}

\subsubsection{RQ2}
As for the similarity of the rendered images compared with the ones obtained by \textsc{\small Stoat}, we use two widely used image similarity metrics~\cite{metrics} (i.e., mean absolute error (MAE) and mean squared error (MSE)) to measure the similarity pixel by pixel. 
MAE measures the average magnitude of differences between the prediction and the actual value and MSE measures the average squared differences between them. 
On average, our rendered images achieve 84\% and 88\% similarity in terms of MAE and MSE, respectively. 
The reasons for the inconsistencies are explained as follows. 
(1) Some data in components are dynamically loaded from web servers or local storage (e.g., \texttt{\small Preference} and SD card), such as the list data of \texttt{\small ListView}, the text data of \texttt{\small TextView}, and the image background of \texttt{\small ImageView}. 
The differences are depicted in Fig.~\ref{fig:render} (a), (b).
(2) Some components (e.g., \texttt{\small Button}) will change their color or visibility after users type some data. For example, as shown in Fig.~\ref{fig:render} (c), the color of the button changes when users input the ``Nickname''. \tool renders the initial state of the activity, thus reducing the similarity of the two UI pages, which however does not affect the understanding of the app functionality.

\subsubsection{RQ3}\label{sec:RQ3}
92 out of 100 activity names are inferred correctly based on our approach, and the accuracy is 92\%. 
To demonstrate and explain the inferring results, we randomly select 10 cases listed in Table~\ref{tbl:labeled}.
The first column refers to the ground-truth activity names.
According to the results of comparison with layout trees in our database, we list the ranks of correct activity names based on the frequency of each name in the second column.
The third column is the corresponding XML layout names. 
\tool accurately infers 8 out of the 10 semantic names which are consistent with the ground truth. 
For \texttt{\small TrackListAct.}, \texttt{\small PersonalInfoAct.} and \texttt{\small SearchAct.} (highlighted in gray color in Table~\ref{tbl:labeled}), the layout names cannot be matched with any name of the candidates, so we choose the name at rank one as the semantic name. 
Two of them do not match the ground truth.
But note that the performance of our approach is highly underestimated, as although some of the recommended names from our \tool differ from the ground truth, they actually have similar meanings, such as \texttt{\small SearchAct.} and \texttt{\small Searcher}.
In addition, with the expansion of our database, the accuracy will be also boosted.
Overall, the results show that \tool can help infer the semantic names of obfuscated activities effectively.

\noindent\fbox{
	\parbox{0.95\linewidth}{
		\textbf{Remark}: \tool outperforms \textsc{\small IC3} on ATG extraction and covers 2 times more activities than \textsc{\small Stoat} with less time. \tool can render UI pages with high similarity (84\%) to the real ones and accurately (92\%) infer the semantic names for obfuscated activity names. 
	}
}

\section{Usefulness Evaluation}
Apart from effectiveness evaluations, we further conduct a user study to demonstrate the usefulness of \tool. Our goals are to check: 
(1) whether \tool can help explore and understand the functionalities of apps effectively?  
(2) whether \tool can help identify the layout code of the given UI page accurately and effectively?

\noindent{\bf Dataset of user study.} 
We randomly select 4 apps (i.e., \texttt{\small Bitcoin}, \texttt{\small Bankdroid}, \texttt{\small ConnectBot}, \texttt{\small Vespucci}) with different number of activities (12-15 activities) from 2 categories (i.e., finance, tool), which are hosted on Google Play Store. 
Each category contains two apps, and we ask participants to explore each app to finish the assigned tasks. 

\noindent{\bf Participant recruitment.} 
We recruit 8 people including postdoc, Ph.D, and master from our university to participate in the experiment via word-of-mouth.
All of the recruited participants have used Android devices for more than one year, and participated in Android related research topics.
They never use these apps before. 
All of them have Android app development experience and come from different countries, such as USA, China, European countries (e.g., Spain), and Singapore. 
The participants receive a \$10 shopping coupon as a compensation of their time.

\noindent{\bf Experiment procedures.} 
We installed the 4 apps on an Android device (Nexus 5 with Android 4.4). 
The experiment started with a brief introduction to the tasks. 
We explained and went through all the features we want them to use within the apps and asked each participant to explore the 4 apps separately to finish the tasks below.
Note that for each category, each participant explored one app with \tool, and the other without \tool. 
To avoid potential bias, the order of app category, and the order of using \tool or not using are rotated based on the Latin Square~\cite{winer1962statistical}.
This setup ensures that each app is explored by multiple participants with different development experience. 
We told each participant to complete two tasks with the given apps: 
(1) manually explore as many functionalities of the apps as possible in 10 minutes, which is far longer than the typical average app session (71.56 seconds)~\cite{bohmer2011falling}, and understand the app functionalities with \tool;
(2) identify the corresponding layout code for the given 2 UI pages in 10 minutes. The 2 UI pages are implemented by static and dynamic layout types respectively.  
After the exploration, 
participants were asked to rate their satisfactoriness in exploration and confidence in mapping UI page and code (on the 5-point likert scale with 1 being least satisfied or confident and 5 being most satisfied or confident).
All participants carried out experiments independently without any discussions with each other.
After performing all tasks, they were required to write some comments about our tool.

\begin{table}
	\scriptsize
	\caption{User study results of app exploration. The figure represents the activity coverage of the 4 apps with manual exploration. $*$ denotes $p$ $<$ 0.01 and $**$ denotes $p$ $<$ 0.05.
	}
\vspace{-1mm}

	\begin{minipage}{0.38\linewidth}
		\centering
		\includegraphics[width=32mm,height=25mm]{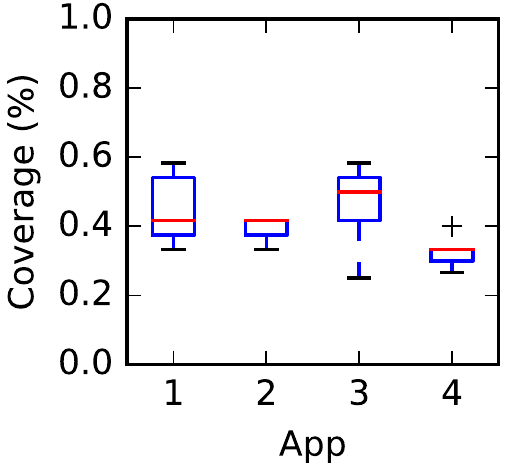}
	\end{minipage}
	\begin{minipage}{0.45\linewidth}
		\vspace{-3mm}
		\begin{tabular}{l c c }
			\toprule
			{\bf Metrics} & \begin{tabular}[c]{@{}c@{}}\textsc{\bf Manual} \\ \textsc{\bf Exploration}\end{tabular} & 
			\begin{tabular}[c]{@{}c@{}}\textsc{\bf Story} \\ -\textsc{\bf Droid}\end{tabular} \\ \toprule
			\begin{tabular}[c]{@{}c@{}}{\bf Time (min)}\end{tabular} & 5.2 & 2.5$^*$ \\
			\begin{tabular}[c]{@{}c@{}}{\bf Coverage}\end{tabular} & 40.8\% & 86.5\%$^*$ \\
			\textsc{\bf Satisfactoriness (1-5)} & 4.2 & 4.4$^*$$^*$ \\ \bottomrule
		\end{tabular}
	\end{minipage}
	\label{tbl:study1}

\end{table}

\noindent{\bf Experiment results.}
As displayed in Table~\ref{tbl:study1}, the average activity coverage of manual exploration is quite low (i.e., 40.8\%), showing the difficulty in exploring app functionalities thoroughly by manual exploration. 
However, the participants' satisfactoriness of completeness of exploration is high (i.e., 4.2 on average).
It indicates that the development teams sometimes are not aware that they miss many features when exploring others' apps.
Such blind confidence and neglection may further negatively influence their strategy or decision in developing their own apps.
Compared with manual exploration, \tool achieves 2 times more activity coverage with less time cost (2.5 minutes on average) to help understand the app functionalities. 
According to the participates' feedback, the average satisfactoriness of \tool is 4.4, which represents the usefulness of helping participants explore and understand app functionalities.  
Table~\ref{tbl:study2} shows the time cost and confidence of mapping UI page to layout code with and without \tool. All the participants spend over 8 minutes to map the UI pages to the corresponding layout code, among which 4 pages are not mapped successfully within 10 minutes. 
In contrast, with the help of \tool, participants only spent 30 seconds to confirm the mapping relations between UI pages and layout code. 
To understand the significance of the differences between without \tool and with \tool, we carry out the Mann-Whitney U test~\cite{utest}, which is designed for small samples.
The results in Table~\ref{tbl:study1} and Table~\ref{tbl:study2} are both significant with $p$-value $<$ 0.01 or $p$-value $<$ 0.05.

We analyze the comments from the participants, and find that they mainly focus on two aspects: (1) It is better to add activity transition methods/events on the edges, such as clicking ``login''; (2) If the transition graph is too complex, \tool needs to provide a better strategy to visualize it.

\section{Discussion}

\subsection{Future Applications Based on \tool}
Apart from the above fundamental usefulness (i.e., exploration of app functionalities), we discuss additional follow-up applications based on the outputs of \tool.

\noindent{\bf Recommendation of UI design and code.} 
Developing the GUI of a mobile application involves two steps, i.e., UI design and implementation. Designing a UI focuses on proper user interaction and visual effects, while implementing a UI focuses on making the UI work as designed with proper layouts and widgets of a GUI framework. 
Our \tool can assist both UI designers and developers by building a large-scale database of app storyboard.
Such a database bridges the gap across the abstract activities (text), UI pages (image) and detailed layout code (i.e., activity$\rightarrow$UI page$\rightarrow$layout code) so that they can be searched as a whole.
Due to that mapping, UI/UX designers can directly use keywords (e.g., ``Login'' and ``Search'') to search the UI images by matching the activity name of the UI in our database.
The searched images can be used for inspiring their own UI design.
The UI developers can also benefit from searching our database for UI implementation.
Given the UI design image from designers, developers can search the similar UI in our database by computing image similarity (e.g., MSE in Section~\ref{sec:RQ3}).
As each UI page in our database is also associated with corresponding run-time UI code, developers can select the most related UI page in the candidate list and then customize the UI code for their own need to implement the given UI design.

\begin{table}[t]
	\caption{User study results of UI page and layout code mapping. $*$ denotes $p$ $<$ 0.01 and $**$ denotes $p$ $<$ 0.05.}
	\vspace{-2mm}
	\scriptsize
	\center
	\label{tbl:study2}
	\begin{tabular}{l c c}
		\toprule
		{\bf Metrics} & {\bf Without \textsc{StoryDroid}} & {\bf With \textsc{StoryDroid}} \\ \toprule
		{\bf Time Cost (min)} & 8.5 & 0.5$^*$ \\
		{\bf Confidence (1-5)} & 4.5 & 5.0$^*$$^*$ \\ \bottomrule
	\end{tabular}
\end{table}

\noindent{\bf Guiding regression testing of apps.}
Reusing test cases is useful to improve the efficiency of regression testing for Android apps~\cite{rothermel2001prioritizing}. \tool can help guide app regression testing by identifying the UI components that have been modified. Different versions of a single app have many common functionalities, which means most of UI pages in the newer version are the same as the previous version. \tool stores the mapping relation between UI page and the corresponding layout code, therefore, analyzers can obtain the modified components by analyzing the differences of layout code and ATG, and further update the corresponding test cases. In this scenario, most of the test cases can be reused, and the modified components can be identified effectively to guide test case update for regression testing.

\subsection{Limitations}
\noindent{\bf Transition Extraction.}
The inputs of UI page rendering are extracted from static analysis based on \textsc{\small Soot}, but some files failed to be transformed, and the call graphs can still be incomplete. As for the closed-source apps, \textsc{\small jadx} is used to decompile apk to Java code. However, some Java files failed to be decompiled, which affects the analysis results of UI page rendering. 
But according to our observation, these cases rarely appear in the real apps.
{Besides, as the activities spawned by other components (e.g., Broadcast Receiver) can only be dynamically loaded, our static-analysis based approach cannot deal with them.}

\noindent{\bf UI Page Rendering.}
(1) Some UI pages use self-defined components by overwriting the method \texttt{\small onDraw}. 
For example, \texttt{\small Kiwix}~\cite{kiwix} lets users read Wikipedia without Internet connections. It draws the canvas for the UI page dynamically. 
(2) Some components are dynamically created from a user-defined function. 
For example, \texttt{\small BankDroid}~\cite{bankdroid} is a banking client for the Swedish banks. It invokes \texttt{\small CreateFrom()} function to create a component where the parameter is the return value of another self-defined method. 
As we cannot get a clue of what these values are unless we analyze them at runtime, we use some empty placeholder to take the position so that the overall layout is the same as the real UI page.

\section{Related Work}
\noindent{\bf Studies for helping Android development.} 
GUI provides a visual bridge between apps and users through which they can interact with each other. 
Developing the GUI of a mobile app involves two separate but related activities: design the UI and implement the UI.
To assist UI implementation, Nguyen and Csallner~\cite{nguyen2015reverse} reverse-engineer the UI screenshots by image processing techniques.
More powerful deep-learning based algorithms~\cite{chen2018ui, beltramelli2018pix2code, moran2018machine} are further proposed to leverage the existing big data of Android apps.
Retrieval-based methods~\cite{reiss2018seeking, behrang2018guifetch} are also used to develop the user interfaces. 
Reiss~\cite{reiss2018seeking} parses the sketch into structured queries to search related UIs of Java-based desktop software in the database.

Different from the UI implementation studies, our study focuses more on the generation of app storyboard which not only contains the UI code, but also the transitions among the UIs.
In addition, the UI code generated in prior work~\cite{nguyen2015reverse, chen2018ui, beltramelli2018pix2code, moran2018machine} is all static layout, which conflicts with our observation in Section~\ref{sec:background} that developers often write Java code to dynamically render the UI. 
In our work, we provide developers with the original UI code (no matter static code, dynamic code, or hybrid) for each screen.
Such real code makes developers more easy to customize the UIs for their own needs.
Apart from the UI implementation, some studies also explore issues between UI design and its implementation.
Moran et al~\cite{moran2018automated} check whether the UI implementation violates the original UI design by comparing the image similarity with computer vision techniques.
They further detect and summarize GUI changes in evolving mobile apps.
They rely on the dynamically running apps for collecting UI screenshots, and that is time-consuming and leads to low coverage of the app.
In contrast, our method can extract most UI pages of the app statically, so it can complement with these studies for related tasks.

\textsc{\small GUIfectch}~\cite{behrang2018guifetch} customizes Reiss's method~\cite{reiss2018seeking} into Android app UI search by considering the transitions between UIs.
It can also extract UI screenshots with corresponding transitions, but our work is different from theirs in two aspects.
First, their model can only deal with open-source apps, while ours can also reverse-engineer the closed-source apps, hence leading to more generality and flexibility.
On the other hand, \textsc{\small GUIfectch} is much more heavy-weight than our static-analysis based approach, as it relies on both static analysis for UI code extraction and dynamic analysis for transition extraction.
In addition, dynamically running the app usually cannot cover all screens, leading to the loss of information.

\noindent{\bf Studies for helping app understanding.} 
The process of reverse engineering of Android apps is that researchers rely on the state-of-the-art tools (e.g., \textsc{\small Apktool}~\cite{apktool}, \textsc{\small Androguard}~\cite{androguard}, \textsc{\small dex2jar}~\cite{dex2jar}, \textsc{\small Soot}~\cite{soot}) for decompiling an \texttt{\small APK} to intermediate language (e.g., \texttt{\small smali}, \texttt{\small jimple}) or Java code. Android reverse engineering is usually used to understand and analyze apps~\cite{understand}. It also can be used to extract features for Android malware detection~\cite{chen2016stormdroid}. However, reverse engineering only has the basic functionality for code review. Different from reverse engineering, our work provides a storyboard of each app to show the basic functionalities and other useful mappings between the UI page and the corresponding layout code, which helps different parties (e.g., PM, UI/UX designer, developer) improve their work efficiency in the real world.

\noindent{\bf Studies for analyzing Android apps.}
Many static analysis techniques~\cite{azim2013targeted,octeau2013effective,octeau2015composite,arzt2014flowdroid,li2015iccta,fan18,fan18efficiently,chen2018mobile,chen2018ausera} have been proposed for Android apps.
\textsc{A$^3$E} provides two strategies, targeted and depth-first exploration, for systematic testing of Android apps~\cite{azim2013targeted}. It also extracts static activity transition graphs for automatically generated test cases. Apart from the target of Android testing, we extract activity transition graphs to identify and systematically explore the storyboard of Android apps. \textsc{\small Epicc} is the first work to extract component communication~\cite{octeau2013effective}, and it determines most Intent attributes to component matching. \textsc{\small ICC}~\cite{octeau2015composite} significantly outperforms \textsc{\small Epicc} on the extraction ability of inter-component communication by utilizing the solver for \texttt{MVC} problems based on the proposed \texttt{COAL} lauguage. 
\textsc{FlowDroid}~\cite{arzt2014flowdroid} and \textsc{IccTA}~\cite{li2015iccta} extract call graphs based on \textsc{\small Soot} for data-flow analysis for detecting data leakage and malicious behaviors~\cite{chen2016stormdroid, fan2016poster,chen2016towards,chen2018automated,chen2019poison,fakeapp}.

\section{Conclusion and Future work}
In this paper, we propose \tool, a system to return visualized storyboard of Android apps by extracting relatively complete ATG, rendering the UI pages statically, and inferring semantic name for obfuscated activities.
Such a storyboard benefits different roles (i.e., PMs, UI designers, and developers) in the app development process. 
The extensive experiment and user study demonstrate the effectiveness and usefulness of \tool. Based on the outputs of \tool, we are able to construct a large-scale database of storyboard to bridge the gap across app activities (text), UI pages (image), and implementation code.
Such a comprehensive database can enable many potential applications such as recommending UI pages to designers and implementation code for developers.
In the future, we will explore these potential applications, and also extend our approach into other platforms such as IOS apps and desktop software for more general usage.
	
\section*{Acknowledgement}
	We appreciate the constructive comments from Prof. Zhenchang Xing and Li Li.
	This work is partially supported by NSFC Grant 61502170, 
	the Science and Technology Commission of Shanghai Municipality Grants 18511103802 and 18511103802, 
	NTU Research Grant NGF-2017-03-033 and NRF Grant CRDCG2017-S04.

\bibliographystyle{IEEEtran}\small
\balance
\bibliography{icse}

\begin{thebibliography}{10}
\providecommand{\url}[1]{#1}
\csname url@samestyle\endcsname
\providecommand{\newblock}{\relax}
\providecommand{\bibinfo}[2]{#2}
\providecommand{\BIBentrySTDinterwordspacing}{\spaceskip=0pt\relax}
\providecommand{\BIBentryALTinterwordstretchfactor}{4}
\providecommand{\BIBentryALTinterwordspacing}{\spaceskip=\fontdimen2\font plus
\BIBentryALTinterwordstretchfactor\fontdimen3\font minus
  \fontdimen4\font\relax}
\providecommand{\BIBforeignlanguage}[2]{{%
\expandafter\ifx\csname l@#1\endcsname\relax
\typeout{** WARNING: IEEEtranS.bst: No hyphenation pattern has been}%
\typeout{** loaded for the language `#1'. Using the pattern for}%
\typeout{** the default language instead.}%
\else
\language=\csname l@#1\endcsname
\fi
#2}}
\providecommand{\BIBdecl}{\relax}
\BIBdecl

\bibitem{understand}
\BIBentryALTinterwordspacing
(2013) How to analyze {APK} and understand it. [Online]. Available:
  \url{https://reverseengineering.stackexchange.com/questions/2703/how-do-i-analyze-a-apk-file-and-understand-its-working}
\BIBentrySTDinterwordspacing

\bibitem{api}
\BIBentryALTinterwordspacing
(2014) Crawl and download apps from {Google} {Play}. [Online]. Available:
  \url{https://github.com/dflower/google-play-crawler}
\BIBentrySTDinterwordspacing

\bibitem{metrics}
\BIBentryALTinterwordspacing
(2016) {Mean Absolute Error} and {Mean Squared Error}. [Online]. Available:
  \url{https://medium.com/human-in-a-machine-world/mae-and-rmse-which-metric-is-better-e60ac3bde13d}
\BIBentrySTDinterwordspacing

\bibitem{typical}
\BIBentryALTinterwordspacing
(2018) 4 steps to develop your app idea. [Online]. Available:
  \url{http://apptology.com/blog/tag/mobile-app-storyboard/}
\BIBentrySTDinterwordspacing

\bibitem{adapter}
\BIBentryALTinterwordspacing
(2018) Android {Adapter}. [Online]. Available:
  \url{https://developer.android.com/reference/android/widget/Adapter}
\BIBentrySTDinterwordspacing

\bibitem{activity}
\BIBentryALTinterwordspacing
(2018) {Android} documentation: Activity. [Online]. Available:
  \url{https://developer.android.com/reference/android/app/Activity}
\BIBentrySTDinterwordspacing

\bibitem{fragment}
\BIBentryALTinterwordspacing
(2018) Android {Fragment}. [Online]. Available:
  \url{https://developer.android.com/reference/android/app/Fragment}
\BIBentrySTDinterwordspacing

\bibitem{naming}
\BIBentryALTinterwordspacing
(2018) {Android} naming conventions. [Online]. Available:
  \url{https://medium.com/@mikelimantara/overview-of-android-project-structure-and-naming-conventions-b08f6d0b7291}
\BIBentrySTDinterwordspacing

\bibitem{packers}
\BIBentryALTinterwordspacing
(2018) {Android Packer Tehchniques}. [Online]. Available:
  \url{http://www.ninoishere.com/android-packer/}
\BIBentrySTDinterwordspacing

\bibitem{bankdroid}
\BIBentryALTinterwordspacing
(2018) Bankdroid. [Online]. Available:
  \url{https://play.google.com/store/apps/details?id=com.liato.bankdroid}
\BIBentrySTDinterwordspacing

\bibitem{web:competitorAnalysis}
\BIBentryALTinterwordspacing
(2018) Competitor analysis before launching a mobile app startup. [Online].
  Available:
  \url{https://growthbug.com/competitor-analysis-before-launching-a-mobile-app-startup-f2f6a19f21b7}
\BIBentrySTDinterwordspacing

\bibitem{csipsimple}
\BIBentryALTinterwordspacing
(2018) Csipsimple. [Online]. Available:
  \url{https://github.com/tqcenglish/CSipSimple}
\BIBentrySTDinterwordspacing

\bibitem{d3}
\BIBentryALTinterwordspacing
(2018) {D3.js}. [Online]. Available: \url{https://d3js.org/}
\BIBentrySTDinterwordspacing

\bibitem{jadx}
\BIBentryALTinterwordspacing
(2018) Dex to {Java} decompiler. [Online]. Available:
  \url{https://github.com/skylot/jadx}
\BIBentrySTDinterwordspacing

\bibitem{fdroid}
\BIBentryALTinterwordspacing
(2018) F-droid {Market}. [Online]. Available:
  \url{https://f-droid.org/en/packages/}
\BIBentrySTDinterwordspacing

\bibitem{web:designer1}
\BIBentryALTinterwordspacing
(2018) Getting better at design is easy, just copy people! [Online]. Available:
  \url{https://medium.com/ux-power-tools/getting-better-at-design-is-easy-just-copy-people-f19ba3be8a62}
\BIBentrySTDinterwordspacing

\bibitem{monkey}
\BIBentryALTinterwordspacing
(2018) {Google Monkey} for {Testing}. [Online]. Available:
  \url{https://developer.android.com/studio/test/monkey}
\BIBentrySTDinterwordspacing

\bibitem{web:appCost}
\BIBentryALTinterwordspacing
(2018) {How Much Does an App Cost}. [Online]. Available:
  \url{https://savvyapps.com/blog/how-much-does-app-cost-massive-review-pricing-budget-considerations}
\BIBentrySTDinterwordspacing

\bibitem{inner}
\BIBentryALTinterwordspacing
(2018) Java {Inner} {Class}. [Online]. Available:
  \url{https://www.tutorialspoint.com/java/java\_innerclasses.htm}
\BIBentrySTDinterwordspacing

\bibitem{kiwix}
\BIBentryALTinterwordspacing
(2018) Kiwix. [Online]. Available:
  \url{https://play.google.com/store/apps/details?id=org.kiwix.kiwixmobile}
\BIBentrySTDinterwordspacing

\bibitem{utest}
\BIBentryALTinterwordspacing
(2018) Mann-{Whitney} {U} test. [Online]. Available:
  \url{http://www.statisticssolutions.com/mann-whitney-u-test/}
\BIBentrySTDinterwordspacing

\bibitem{devprocess}
\BIBentryALTinterwordspacing
(2018) Mobile app development process. [Online]. Available:
  \url{https://thebhwgroup.com/blog/mobile-app-development-process}
\BIBentrySTDinterwordspacing

\bibitem{web:mobileDesktop}
\BIBentryALTinterwordspacing
(2018) Mobile {Internet} use passes desktop for the first time. [Online].
  Available:
  \url{https://techcrunch.com/2016/11/01/mobile-internet-use-passes-desktop-for-the-first-time-study-finds/}
\BIBentrySTDinterwordspacing

\bibitem{web:appNumber}
\BIBentryALTinterwordspacing
(2018) Number of apps available in leading app stores as of 1st quarter.
  [Online]. Available:
  \url{https://www.statista.com/statistics/276623/number-of-apps-available-in-leading-app-stores/}
\BIBentrySTDinterwordspacing

\bibitem{storydroid}
\BIBentryALTinterwordspacing
(2018) Overview of {StoryDroid}. [Online]. Available:
  \url{https://sites.google.com/view/storydroid/}
\BIBentrySTDinterwordspacing

\bibitem{podlisten}
\BIBentryALTinterwordspacing
(2018) Podlisten. [Online]. Available:
  \url{https://f-droid.org/en/packages/com.einmalfel.podlisten/}
\BIBentrySTDinterwordspacing

\bibitem{androguard}
\BIBentryALTinterwordspacing
(2018) Reverse engineering of {Android} applications. [Online]. Available:
  \url{https://github.com/androguard/androguard}
\BIBentrySTDinterwordspacing

\bibitem{soot}
\BIBentryALTinterwordspacing
(2018) Soot: A {Java} optimization framework. [Online]. Available:
  \url{https://github.com/Sable/soot}
\BIBentrySTDinterwordspacing

\bibitem{apktool}
\BIBentryALTinterwordspacing
(2018) A tool for reverse engineering {Android} apk files. [Online]. Available:
  \url{https://ibotpeaches.github.io/Apktool/}
\BIBentrySTDinterwordspacing

\bibitem{dex2jar}
\BIBentryALTinterwordspacing
(2018) Tools to work with {Android} .dex and {Java} .class files. [Online].
  Available: \url{https://github.com/pxb1988/dex2jar}
\BIBentrySTDinterwordspacing

\bibitem{twitter}
\BIBentryALTinterwordspacing
(2018) Twitter. [Online]. Available:
  \url{https://play.google.com/store/apps/details?id=com.twitter.android}
\BIBentrySTDinterwordspacing

\bibitem{web:designer2}
\BIBentryALTinterwordspacing
(2018) Uninvited {Redesigns}. [Online]. Available:
  \url{https://uninvitedredesigns.com/}
\BIBentrySTDinterwordspacing

\bibitem{vespucci}
\BIBentryALTinterwordspacing
(2018) Vespucci. [Online]. Available:
  \url{https://play.google.com/store/apps/details?id=de.blau.android}
\BIBentrySTDinterwordspacing

\bibitem{arbon2014app}
J.~J. Arbon, \emph{App quality: Secrets for agile app teams}.\hskip 1em plus
  0.5em minus 0.4em\relax Jason Arbon, 2014.

\bibitem{arzt2014flowdroid}
S.~Arzt, S.~Rasthofer, C.~Fritz, E.~Bodden, A.~Bartel, J.~Klein, Y.~Le~Traon,
  D.~Octeau, and P.~McDaniel, ``Flowdroid: {Precise} context, flow, field,
  object-sensitive and lifecycle-aware taint analysis for {Android} apps,''
  \emph{Acm Sigplan Notices}, vol.~49, no.~6, pp. 259--269, 2014.

\bibitem{azim2013targeted}
T.~Azim and I.~Neamtiu, ``Targeted and depth-first exploration for systematic
  testing of {Android} apps,'' in \emph{Acm Sigplan Notices}, vol.~48,
  no.~10.\hskip 1em plus 0.5em minus 0.4em\relax ACM, 2013, pp. 641--660.

\bibitem{behrang2018guifetch}
F.~Behrang, S.~P. Reiss, and A.~Orso, ``Guifetch: {Supporting} app design and
  development through {GUI} search,'' in \emph{Proceedings of the 5th
  International Conference on Mobile Software Engineering and Systems}.\hskip
  1em plus 0.5em minus 0.4em\relax ACM, 2018, pp. 236--246.

\bibitem{beltramelli2018pix2code}
T.~Beltramelli, ``pix2code: {Generating} code from a graphical user interface
  screenshot,'' in \emph{Proceedings of the ACM SIGCHI Symposium on Engineering
  Interactive Computing Systems}.\hskip 1em plus 0.5em minus 0.4em\relax ACM,
  2018, p.~3.

\bibitem{bohmer2011falling}
M.~B{\"o}hmer, B.~Hecht, J.~Sch{\"o}ning, A.~Kr{\"u}ger, and G.~Bauer,
  ``Falling asleep with {Angry Birds}, {Facebook} and {Kindle}: a large scale
  study on mobile application usage,'' in \emph{Proceedings of the 13th
  international conference on Human computer interaction with mobile devices
  and services}.\hskip 1em plus 0.5em minus 0.4em\relax ACM, 2011, pp. 47--56.

\bibitem{chen2018ui}
C.~Chen, T.~Su, G.~Meng, Z.~Xing, and Y.~Liu, ``From {UI} design image to {GUI}
  skeleton: {A} neural machine translator to bootstrap mobile {GUI}
  implementation,'' in \emph{Proceedings of the 40th International Conference
  on Software Engineering}.\hskip 1em plus 0.5em minus 0.4em\relax ACM, 2018,
  pp. 665--676.

\bibitem{chen2019codegeneration}
S.~Chen, L.~Fan, T.~Su, L.~Ma, Y.~Liu, and L.~Xu, ``Automated cross-platform
  {GUI} code generation for mobile apps,'' in \emph{Proceedings of the 26th
  IEEE International Conference on Software Analysis, Evolution and
  Reengineering, {SANER}}.\hskip 1em plus 0.5em minus 0.4em\relax IEEE, 2019.

\bibitem{chen2018ausera}
S.~Chen, G.~Meng, T.~Su, L.~Fan, Y.~Xue, Y.~Liu, L.~Xu, M.~Xue, B.~Li, and
  S.~Hao, ``{AUSERA}: Large-scale automated security risk assessment of global
  mobile banking apps,'' \emph{arXiv preprint arXiv:1805.05236}, 2018.

\bibitem{chen2018mobile}
S.~Chen, T.~Su, L.~Fan, G.~Meng, M.~Xue, Y.~Liu, and L.~Xu, ``Are mobile
  banking apps secure? {What} can be improved?'' in \emph{Proceedings of the
  2018 26th ACM Joint Meeting on European Software Engineering Conference and
  Symposium on the Foundations of Software Engineering}.\hskip 1em plus 0.5em
  minus 0.4em\relax ACM, 2018, pp. 797--802.

\bibitem{chen2018automated}
S.~Chen, M.~Xue, L.~Fan, S.~Hao, L.~Xu, H.~Zhu, and B.~Li, ``Automated
  poisoning attacks and defenses in malware detection systems: {An} adversarial
  machine learning approach,'' \emph{computers \& security}, vol.~73, pp.
  326--344, 2018.

\bibitem{chen2019poison}
S.~Chen, M.~Xue, L.~Fan, L.~Ma, Y.~Liu, and L.~Xu, ``How can we craft
  large-scale {Android} malware? {An} automated poisoning attack,'' in
  \emph{Proceedings of the 26th IEEE International Conference on Software
  Analysis, Evolution and Reengineering, {SANER}}.\hskip 1em plus 0.5em minus
  0.4em\relax IEEE, 2019.

\bibitem{chen2016stormdroid}
S.~Chen, M.~Xue, Z.~Tang, L.~Xu, and H.~Zhu, ``Stormdroid: A streaminglized
  machine learning-based system for detecting {Android} malware,'' in
  \emph{Proceedings of the 11th ACM on Asia Conference on Computer and
  Communications Security, {ASIACCS}}.\hskip 1em plus 0.5em minus 0.4em\relax
  ACM, 2016, pp. 377--388.

\bibitem{chen2016towards}
S.~Chen, M.~Xue, and L.~Xu, ``Towards adversarial detection of mobile malware:
  poster,'' in \emph{Proceedings of the 22nd Annual International Conference on
  Mobile Computing and Networking}.\hskip 1em plus 0.5em minus 0.4em\relax ACM,
  2016, pp. 415--416.

\bibitem{dit2011can}
B.~Dit, L.~Guerrouj, D.~Poshyvanyk, and G.~Antoniol, ``Can better identifier
  splitting techniques help feature location?'' in \emph{2011 19th IEEE
  International Conference on Program Comprehension}.\hskip 1em plus 0.5em
  minus 0.4em\relax IEEE, 2011, pp. 11--20.

\bibitem{dong2018understanding}
S.~Dong, M.~Li, W.~Diao, X.~Liu, J.~Liu, Z.~Li, F.~Xu, K.~Chen, X.~Wang, and
  K.~Zhang, ``Understanding {Android} obfuscation techniques: {A} large-scale
  investigation in the wild,'' \emph{arXiv preprint arXiv:1801.01633}, 2018.

\bibitem{fan18efficiently}
L.~Fan, T.~Su, S.~Chen, G.~Meng, Y.~Liu, L.~Xu, and G.~Pu, ``Efficiently
  manifesting asynchronous programming errors in {Android} apps,'' in
  \emph{Proceedings of the 2018 33rd ACM/IEEE International Conference on
  Automated Software Engineering, {ASE}, Montpellier, France, May 27 - June
  03}, 2018, pp. 485--496.

\bibitem{fan18}
L.~Fan, T.~Su, S.~Chen, G.~Meng, Y.~Liu, L.~Xu, G.~Pu, and Z.~Su, ``Large-scale
  analysis of framework-specific exceptions in {Android} apps,'' in
  \emph{Proceedings of the 40th International Conference on Software
  Engineering, {ICSE} 2018, Gothenburg, Sweden, May 27 - June 03, 2018}, 2018,
  pp. 408--419.

\bibitem{fan2016poster}
L.~Fan, M.~Xue, S.~Chen, L.~Xu, and H.~Zhu, ``Poster: Accuracy vs. time cost:
  Detecting {Android} malware through pareto ensemble pruning,'' in
  \emph{Proceedings of the 2016 ACM SIGSAC Conference on Computer and
  Communications Security}.\hskip 1em plus 0.5em minus 0.4em\relax ACM, 2016,
  pp. 1748--1750.

\bibitem{finch1995art}
C.~Finch and P.~Blake, \emph{The art of Walt Disney: From Mickey mouse to the
  magic kingdoms}.\hskip 1em plus 0.5em minus 0.4em\relax Abrams, 1995.

\bibitem{fox2017mobile}
R.~Fox, ``Mobile app development: The effect of smartphones, mobile
  applications and geolocation services on the tourist experience,'' Ph.D.
  dissertation, University of Baltimore, 2017.

\bibitem{guo2017automated}
L.~Guo, R.~Sharma, L.~Yin, R.~Lu, and K.~Rong, ``Automated competitor analysis
  using big data analytics: Evidence from the fitness mobile app business,''
  \emph{Business Process Management Journal}, vol.~23, no.~3, pp. 735--762,
  2017.

\bibitem{li2015iccta}
L.~Li, A.~Bartel, T.~F. Bissyand{\'e}, J.~Klein, Y.~Le~Traon, S.~Arzt,
  S.~Rasthofer, E.~Bodden, D.~Octeau, and P.~McDaniel, ``Iccta: {Detectingz}
  inter-component privacy leaks in {Android} apps,'' in \emph{Proceedings of
  the 37th International Conference on Software Engineering-Volume 1}.\hskip
  1em plus 0.5em minus 0.4em\relax IEEE Press, 2015, pp. 280--291.

\bibitem{li2014know}
L.~Li, A.~Bartel, J.~Klein, Y.~L. Traon, S.~Arzt, S.~Rasthofer, E.~Bodden,
  D.~Octeau, and P.~Mcdaniel, ``I know what leaked in your pocket: {Uncovering}
  privacy leaks on {Android} apps with static taint analysis,'' \emph{arXiv
  preprint arXiv:1404.7431}, 2014.

\bibitem{sapienz}
K.~Mao, M.~Harman, and Y.~Jia, ``Sapienz: {Multi-objective} automated testing
  for {Android} applications,'' in \emph{Proceedings of the 25th International
  Symposium on Software Testing and Analysis}.\hskip 1em plus 0.5em minus
  0.4em\relax ACM, 2016, pp. 94--105.

\bibitem{moran2018machine}
K.~Moran, C.~Bernal-C{\'a}rdenas, M.~Curcio, R.~Bonett, and D.~Poshyvanyk,
  ``Machine learning-based prototyping of graphical user interfaces for mobile
  apps,'' \emph{arXiv preprint arXiv:1802.02312}, 2018.

\bibitem{moran2018automated}
K.~Moran, B.~Li, C.~Bernal-C{\'a}rdenas, D.~Jelf, and D.~Poshyvanyk,
  ``Automated reporting of {GUI} design violations for mobile apps,''
  \emph{arXiv preprint arXiv:1802.04732}, 2018.

\bibitem{nguyen2015reverse}
T.~A. Nguyen and C.~Csallner, ``Reverse engineering mobile application user
  interfaces with remaui (t),'' in \emph{Automated Software Engineering (ASE),
  2015 30th IEEE/ACM International Conference on}.\hskip 1em plus 0.5em minus
  0.4em\relax IEEE, 2015, pp. 248--259.

\bibitem{octeau2015composite}
D.~Octeau, D.~Luchaup, M.~Dering, S.~Jha, and P.~McDaniel, ``Composite constant
  propagation: Application to {Android} inter-component communication
  analysis,'' in \emph{Proceedings of the 37th International Conference on
  Software Engineering-Volume 1}.\hskip 1em plus 0.5em minus 0.4em\relax IEEE
  Press, 2015, pp. 77--88.

\bibitem{octeau2013effective}
D.~Octeau, P.~McDaniel, S.~Jha, A.~Bartel, E.~Bodden, J.~Klein, and
  Y.~Le~Traon, ``Effective inter-component communication mapping in {Android}
  with epicc: An essential step towards holistic security analysis,''
  \emph{Effective Inter-Component Communication Mapping in {Android} with
  Epicc: An Essential Step Towards Holistic Security Analysis}, 2013.

\bibitem{reiss2018seeking}
S.~P. Reiss, Y.~Miao, and Q.~Xin, ``Seeking the user interface,''
  \emph{Automated Software Engineering}, pp. 157--193, 2018.

\bibitem{rothermel2001prioritizing}
G.~Rothermel, R.~H. Untch, C.~Chu, and M.~J. Harrold, ``Prioritizing test cases
  for regression testing,'' \emph{IEEE Transactions on software engineering},
  vol.~27, no.~10, pp. 929--948, 2001.

\bibitem{su2016fsmdroid}
T.~Su, ``Fsmdroid: guided {GUI} testing of {Android} apps,'' in \emph{IEEE/ACM
  International Conference on Software Engineering Companion (ICSE-C)}.\hskip
  1em plus 0.5em minus 0.4em\relax IEEE, 2016, pp. 689--691.

\bibitem{su2017guided}
T.~Su, G.~Meng, Y.~Chen, K.~Wu, W.~Yang, Y.~Yao, G.~Pu, Y.~Liu, and Z.~Su,
  ``Guided, stochastic model-based {GUI} testing of {Android} apps,'' in
  \emph{Proceedings of the 2017 11th Joint Meeting on Foundations of Software
  Engineering}.\hskip 1em plus 0.5em minus 0.4em\relax ACM, 2017.

\bibitem{fakeapp}
C.~Tang, S.~Chen, L.~Fan, L.~Xu, Y.~Liu, Z.~Tang, and L.~Dou, ``A large-scale
  empirical study on industrial fake apps,'' in \emph{Proceedings of the 41th
  ACM/IEEE International Conference on Software Engineering, {ICSE}}.\hskip 1em
  plus 0.5em minus 0.4em\relax IEEE, 2019.

\bibitem{winer1962statistical}
B.~J. Winer, ``Statistical principles in experimental design.'' 1962.

\bibitem{zhang1989simple}
K.~Zhang and D.~Shasha, ``Simple fast algorithms for the editing distance
  between trees and related problems,'' \emph{SIAM journal on computing},
  vol.~18, no.~6, pp. 1245--1262, 1989.

\end{thebibliography}
	
\end{document}